\documentclass[preprint]{aastex}

\usepackage{graphicx}
\usepackage{natbib}
\bibpunct{(}{)}{;}{a}{}{,}

\newcommand{\othree}{[O~\footnotesize{III}\normalsize]}

\newcommand{\ntwo}{[N~\footnotesize{II}\normalsize]}

\begin{document}

\title{Minor Galaxy Interactions: Star Formation Rates and Galaxy Properties}

\author{Deborah Freedman Woods}
\affil{Department of Astronomy, Harvard University
\\ 60 Garden St., Cambridge, MA 02138}
\email{dwoods@cfa.harvard.edu}

\author{Margaret J. Geller}
\affil{Smithsonian Astrophysical Observatory
\\ 60 Garden St., Cambridge, MA 02138}

\begin{abstract}

We study star formation in a sample of 1204 galaxies in minor 
($\left | \Delta m_z \right | \geq 2$) pairs and compact groups, drawn 
from the Sloan Digital Sky Survey Data Release 5 (SDSS DR5).  We analyze an 
analogous sample of 2409 galaxies in major ($\left | \Delta m_z \right | < 2$) 
pairs and compact groups to ensure that our selection reproduces previous
results, and we use a ``field'' sample of  65,570 galaxies for comparison.  
Our major and minor pairs samples include only galaxies 
in spectroscopically confirmed pairs, where the recessional velocity separation 
$\Delta V < 500$~km s$^{-1}$ and the projected spatial separation 
$\Delta D < 50$~kpc h$^{-1}$.  
The relative magnitude (a proxy for the mass ratio) of the pair 
is an important parameter in the effectiveness of the tidally triggered star 
formation in minor interactions.  As expected, the secondary galaxies in
minor pairs show evidence for tidally triggered star formation, whereas the
primary galaxies in the minor pairs do not.  The galaxy color is also an 
important parameter in the effectiveness of triggered star  formation in the 
major galaxy pairs. In the major pairs sample, there is a correlation between
the specific H$\alpha$ star formation rate (SSFR) and $\Delta D$ in the blue
primary and blue secondary galaxies; for the red primary and red secondary 
galaxies, there is none.  Galaxies in pairs have a higher mean SSFR at every
absolute magnitude compared to matched sets of field galaxies, and the 
relative increase in mean SSFR  becomes larger with decreasing intrinsic 
luminosity.  We also detect a significantly increased AGN fraction  
 in the pair galaxies compared to matched sets of field galaxies.

\end{abstract}


\section{Introduction}
\setcounter{footnote}{1}

Observations of apparently interacting galaxies in nearby systems 
reveal enhanced star formation compared to non-interacting
galaxies.  \citet{larson_tinsley} first showed that systems in the
Atlas of Peculiar Galaxies \citep{arp66} have a wider range of 
optical colors and star formation rates than typical galaxies.
Observations of increased star formation activity in galaxy interactions
include measurements of H$\alpha$ and other optical emission lines  
\citep{kk84,kenn87,keel93,liu-kenn95a,donzelli-p97,bgk,bgk03,lambas03,nikolic04,kauf04},
infrared emission \citep{kenn87,jonesstein89,seg-wol92,keel93,nikolic04,geller06}, 
radio continuum emission \citep{hummel81}, and galaxy colors 
\citep{larson_tinsley,geller06}.  \citet{struck05} reviews
galaxy collisions.

Numerical simulations of major galaxy interactions account for the general features
of the observations.  The simulations of \citet{toomre72} first demonstrated 
 that gravitational interactions between galaxies produce
disrupted morphologies like those observed.  Other early simulations describe 
how galaxy mergers deposit gas in the centers of merger remnants 
\citep{negroponte83,hernquist89,barnes+hern91}.  
Central star formation in interacting galaxies results from gaseous inflows
that occur when the gas loses angular momentum through
 gravitational  tidal torques produced primarily by the 
non-axisymmetric structure induced by the companion galaxy \citep{mihos+hern96}.  
The galaxy structure strongly influences the strength
and timing of the burst of star formation triggered by the gaseous inflows
\citep[e.g.][]{mihos+hern96,tissera}.
The simulations of \citet{cox06} model colliding disk 
galaxies with feedback from massive stars and radiative cooling, and use a 
density-dependent star formation prescription; they produce 
a centrally concentrated burst of star formation for simulations with a 
range of star formation histories.  (See the introduction in \citet{cox06} 
for a concise summary of galaxy merger simulations.)

Simulations of the Milky Way Galaxy and Large Magellanic Cloud (LMC)
interaction suggest that there should be an asymmetry in the response to the 
gravitational interaction of the primary and secondary galaxies in  minor 
interactions\footnotemark[1].  \citet{mastro} show in their simulations 
of the Milky Way Galaxy and LMC interaction that the LMC experiences tidal 
disruption causing an elongated disc, bar creation, warped profile, diffuse 
stellar halo, and gas loss from the LMC's disc.  Meanwhile, the Milky Way 
disc shows essentially no response to the encounter.
The simulations of \citet{mayer01} of the Milky Way and dwarf irregular 
galaxies reveal central infall of gas and gas stripping in the dwarf irregular 
galaxies.  The high central surface brightness dwarf galaxies experience a single, 
strong burst of central starformation; the low central surface brightness dwarf
galaxies experience multiple smaller epochs of star formation. 
There is, however, a plausible mechanism for induced star formation in the major 
galaxies: simulations show that tidal torques from minor companions provoke 
non-axisymmetric structure in the main disk galaxy, causing the gas to 
lose its angular momentum and fall inward, presumably leading to
central star formation \citep{hern+mihos95}.

\footnotetext{To avoid potentially confusing language, we refer to galaxy 
pairs with magnitude difference $\left | \Delta m \right | <2$ as  ``major'' 
pairs, and those with $\left | \Delta m \right | \geq 2$ as ``minor'' pairs.
In both cases, we refer to the more luminous galaxy in the pair as the 
``primary'' galaxy, and the less luminous as the ``secondary'' galaxy. }

There is debate over the connection between galaxy interactions and enhanced 
AGN activity.  A number of studies investigate the
enhanced AGN fraction in interacting systems 
\citep{dahari85,keel85,bushouse,kelm,alonso07} 
and the possible excess of companions to Seyfert galaxies 
\citep{fuentes88,schmitt01}.  Although the results have not 
always agreed, there is growing observational evidence for
interaction-driven nuclear activity \citep[e.g.][]{hennawi06,serber06}.

Large data sets enable statistical analysis of the relationship
between star formation and pair properties.  In their sample of $\sim 500$ 
galaxies drawn from the CfA2 Redshift Survey, \citet{bgk} show that the 
equivalent width of the H$\alpha$ emission (EW(H$\alpha$)) is anti-correlated 
with the projected galaxy pair separation ($\Delta D$).  Numerous additional 
studies confirm the anti-correlation  between star formation indicators 
and projected separation in the SDSS \citep{nikolic04}, the 2dF Survey 
\citep{lambas03}, and the CfA2 Redshift Survey plus infrared data 
\citep{geller06}.  The galaxy pairs identified in these data sets are 
dominated by major interactions, where the magnitude differences 
in the systems are small.
This property of pair samples is a natural consequence of the magnitude 
distribution in magnitude-limited redshift surveys; pairs with large magnitude 
difference reside in the tails of the distribution.  Targeted searches for high 
contrast pairs have inherently low success rates because the background galaxies 
outnumber the probable companions.

Although difficult to identify, minor companions are particularly interesting 
because they play a critical role in the formation and evolution of galaxies
\citep{somerville_prim99,kauffmann99a,kauffmann99b,diaferio99}.  
  Hierarchical structure formation 
models show that galaxies grow by accreting other galaxies, most
often minor companions (see the merger tree in \citealt{wechsler02}).  
Minor mergers may be at least an order of magnitude more common
than major mergers over the course of history (see 
\citet{hern+mihos95} and references therein).  Despite their role in 
galaxy  formation models, observational studies of minor interactions are
rare.  A study of minor interactions from the CfA2 Redshift Survey and
follow-up observations by \citet{woods} includes 57
galaxies with magnitude difference ($\left | \Delta m_r \right | > 2$).
Woods et al. find that  star formation activity in their sample of 
minor interactions is  not enhanced as strongly as it is in major interactions;
however, their small data set precludes analyzing the primary and 
secondary galaxies separately.

Here we address the effectiveness of minor interactions in triggering 
star formation.  We examine the way the effectiveness differs in the primary and 
secondary galaxies.  We also study how intrinsic galaxy properties 
including luminosity and color influence the triggered star formation in both 
major and minor interactions.  We also investigate the preferential
association of AGNs with interacting systems.  We assume the standard flat
$\Lambda CDM$ cosmological model with a Hubble constant of 
$H_0 = 71$~km~s$^{-1}$~Mpc$^{-1}$ and $\Omega_m h^2= 0.135$ \citep{spergel}, 
and we use the convention $H_0 = 100 h$.

In \S\ref{selection}, 
we discuss the selection of the major and minor pairs samples, and of
the field galaxy sample used for comparison.  We describe the samples'
properties in \S\ref{properties}, including absolute and relative magnitudes, 
and colors.  We discuss the relative abundance of AGN in various samples 
and their relation to pair properties in \S\ref{agns}.  The star formation 
rates in our pairs samples and field sample are discussed in 
\S\ref{sfr}.  We compare the star formation rates with pair 
properties in \S\ref{sfrprop} and with galaxy properties in \S\ref{sfrmag}. 
We conclude in \S\ref{conclusion}.

\section{Sample Selection} \label{selection}

\subsection{Galaxy Pairs Selection} \label{pairssample}
\setcounter{footnote}{1}

Although minor interactions must occur frequently in the
galaxy formation process \citep[e.g.][]{wechsler02}, they are not 
well observed.  Identifying minor companions observationally is challenging 
because magnitude limited redshift surveys naturally contain relatively 
more pairs of similar magnitude, and because directed searches for 
low-luminosity companions around primary galaxies have inherently low 
success rates due to contamination by the more abundant background 
galaxies \citep{woods}.  The frequency of false pairs, especially 
for the case of minor companions, makes it essential to have spectroscopic
confirmation of coincidence in redshift space.  

The sheer size of the Sloan Digital Sky Survey Data Release 5 (SDSS DR5) with 
more than 8,000 square degrees of the sky observed and spectra for 674,749 
galaxies to $m_r = 17.77$ allows us to find a significant sample of high
contrast pairs\footnotemark[1].  \citet{york} contains a technical summary of SDSS;
\citet{gunn} describe the camera; \citet{telescope} describe the telescope;
\citet{fukugita96} has the sdss filter definitions; \citet{strauss}
introduces the Main Galaxy Sample; and \cite{sdss} contains a description of 
DR4 (no DR5 summary paper exists to date).  

\footnotetext{Sloan Digital Sky Survey.  2007, SDSS  Website: 
http://www.sdss.org/dr5/}

We select our pair sample with projected physical separation 
$\Delta D < 50$~kpc h$^{-1}$ and line-of-sight velocity separation 
$\Delta V < 500$~km s$^{-1}$ from the nearest neighbor.   To 
minimize aperture effects, we restrict our sample to the redshift 
range $8,000 < cz < 50,000$~km s$^{-1}$, and we require that at least 20\% 
of the total galaxy luminosity lands in the fiber.  \citet{kewley05}
demonstrate that a covering fraction of $> 20\%$ is necessary to avoid 
substantial differences in nuclear and global measurements of star formation
rates, extinction, and metallicity in their sample of 101 galaxies from the
Nearby Field Galaxy Survey \citep{nfgs1,nfgs2}.  The main effect
of this criterion is to exclude very extended nearby galaxies.  Although
this selection introduces a slight bias toward centrally concentrated galaxies at
low redshift, that bias is the same for the pair and field galaxy samples.

We further refine our sample to avoid rich clusters and to reduce the 
influence from other neighboring galaxies.  To minimize effects of the 
morphology-density relation \citep{dressler} or the SFR-density
relation \citep{gomez03,lewis02,hashimoto98}, we exclude
galaxies within $\Delta D < 1.5$~Mpc h$^{-1}$ ($\sim 1$ virial radius)
and $\Delta V < 5000$~km s$^{-1}$ 
of rich Abell clusters.  \citet{rines05} show in their Cluster and Infall
Region Nearby Survey (CAIRNS) that outside 2-3 virial radii from the 
cluster center, the fraction of galaxies with H$\alpha$ star formation 
detected in their inner disks is comparable to that of the field.
Moreover, the distribution of star formation rates is independent of 
global density \citep{rines05,carter01}.  In the context of galaxy
interactions, the gravitational tidal interaction can potentially trigger
enhanced star formation activity whenever the galaxy contains enough gas.  
Galaxies not in clusters are more likely to have substantial gas available
for star formation, providing a stronger signal of the interaction.

After constructing the initial pairs catalog, we further
exclude galaxies that have another apparent 
photometric companion ($\left | \Delta m_z \right | < 2$, compared to the
primary) without a measured redshift and located at smaller 
projected separation than the spectroscopically confirmed companion.  
Otherwise, in our measurements of $\Delta D$ and $\left | \Delta m_z \right |$,
it would be unclear which galaxy is actually the nearest projected neighbor.
Because we compare each galaxy to its nearest neighbor, there are 
situations where we include only partial pairs in the final analysis.
Only one member of an apparent group may meet our
requirement that no closest companion lacks spectroscopy; other
 members of the group may fail the test.  We also accept partial pairs
in the case where  one member of the pair fails to meet our aperture test
 or has a poor quality spectrum, but has a reliable magnitude and redshift 
reported allowing measurement of $\Delta D$ and $\left | \Delta m_z \right |$.

The spectroscopic data for our sample, including the spectral line 
fits, come from the Princeton reductions of the SDSS DR5.  The photometric 
data are provided by the SDSS DR5.  We take our data from local copies 
of the SDSS and Princeton datasets, maintained at the Harvard-Smithsonian 
Center for Astrophysics.  We select the sample in SDSS Petrosian 
$z$-band for the best correspondence between luminosity and mass, and we
divide it into subsets of major and minor interactions.  

\citet{nikolic04} show that SDSS $z$-band
Petrosian magnitudes provide an unbiased estimator of galaxy mass
in SDSS pairs by comparing the $m_z$ with total $K$-band magnitudes 
from the Two-Micron All-Sky Survey (2MASS) Extended Source Catalog 
(XSC; \citealp{jarrett}), which are nearly independent of extinction.
The primary galaxies in our sample have $m_z < 17.77$, which
is a subset of the galaxies within the SDSS spectroscopy completeness 
limit of $m_r < 17.77$, but provides consistency when using $z$-band 
photometry.  All of the magnitudes quoted in this paper are Petrosian 
\citep{petrosian,strauss}.  

We use the H$\alpha$ emission line to calculate the star formation
rate of our pair and field galaxies.  We require that H$\alpha$ and 
H$\beta$ flux have signal-to-noise ratio $>3$ for accurate
measurement of specific H$\alpha$ star formation rates (SSFRs) using 
optical data.  Numerous studies demonstrate an anti-correlation between star 
formation indicators and projected separation, using a variety of star formation 
indicators.  \citet{bgk,bgk03} and \citet{woods} use EW(H$\alpha$).  
The EW measurement is independent of reddening, unlike the H$\alpha$ star 
formation rate.  A complete census of interacting galaxies, including 
highly reddened objects, requires infrared data \citep{geller06}.  The 
galaxy pairs in Geller et al. show an anti-correlation between star 
formation rate and $\Delta D$, for both far-IR and EW(H$\alpha$) star 
formation indicators, as do the galaxy pairs in \citet{nikolic04}.  

Visual inspection of all of the pair galaxies allows us to eliminate
some false pairs from our sample.  The main problem we find is the
automated deblending of a single extended galaxy into multiple galaxies, each 
with different photometric and spectroscopic measurements.   We estimate that
$\sim15\%$ of the minor pairs with projected separation $\Delta D < 50$~kpc h$^{-1}$ 
are actually a poorly deblended galaxy.  This problem is most common for 
potential pairs at small projected separation.   We choose not to exclude 
all galaxies that are flagged by SDSS as deblended because doing so would eliminate 
substantial numbers of real pairs at small projected separation, which are 
important for our sample.  Failure to inspect the close pairs, however, could 
be problematic for galaxy counts and other measurements.

Although this paper focuses on the star formation properties of the minor
galaxy pairs, we include the major galaxy pairs to make sure that our
selection reproduces previous results.  Major galaxy pairs have been well 
studied, but the difference in sample selection can make it difficult to draw 
direct comparisons between minor and major pairs samples.  By 
constructing a sample of major companions with exactly the same
selection effects as the minor companions, we can draw a better
comparison between the two sets of pairs.

The sample of minor mergers, defined here as 
$\left | \Delta m_z \right | \geq 2$, comprises 1204 galaxies in
pairs or compact groups. The sample of major mergers, defined here as 
$\left | \Delta m_z \right | < 2$, has 2409 galaxies in 
pairs and compact groups.  A magnitude difference 
$\left | \Delta m_z \right | = 2$ corresponds to a mass ratio of
$\sim6$, assuming a constant mass-to-light ratio across the $M_z$ range.

\begin{figure}[htb!]
\centering
  \includegraphics[width=2.3in,angle=90]{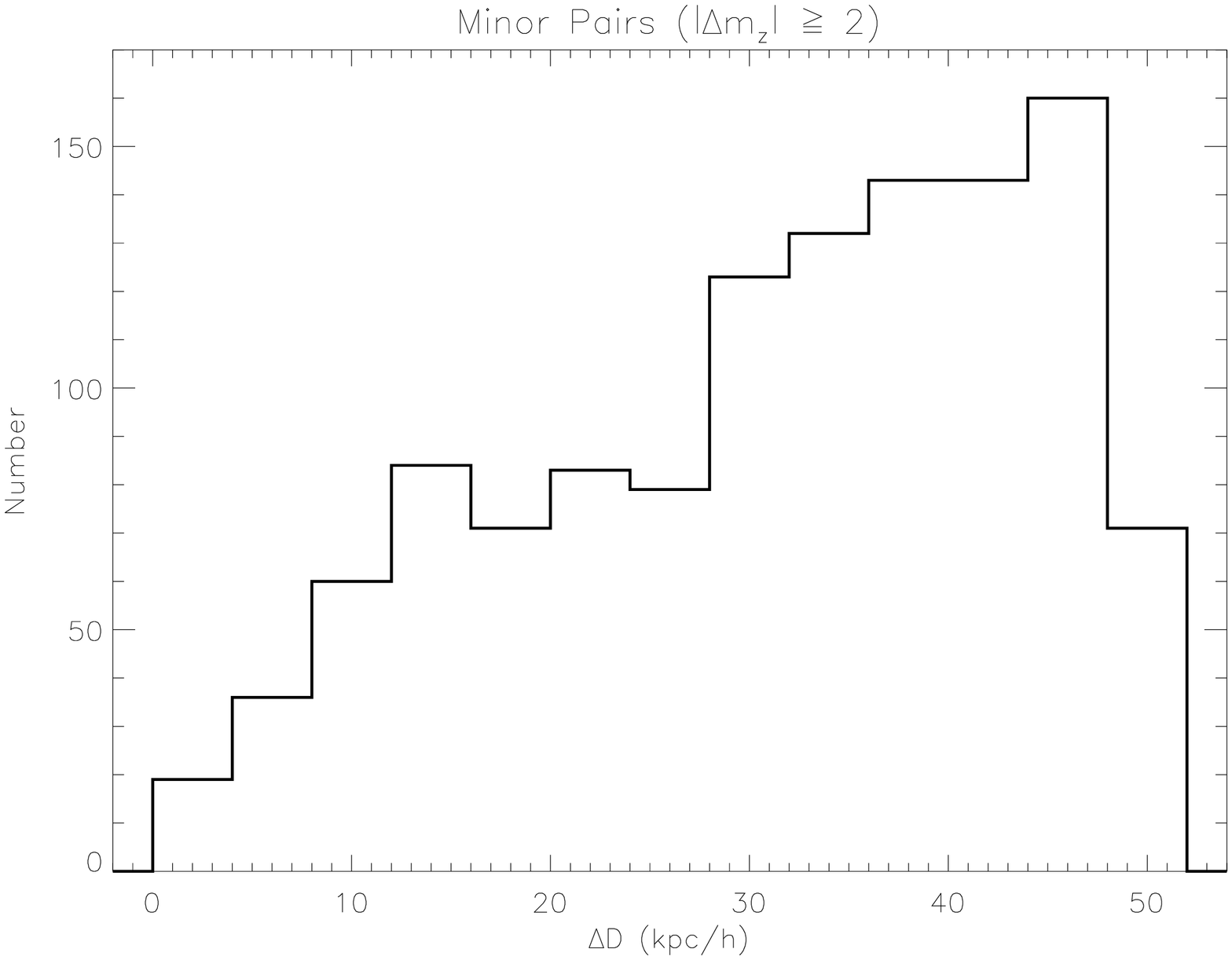}
 \includegraphics[width=2.3in,angle=90]{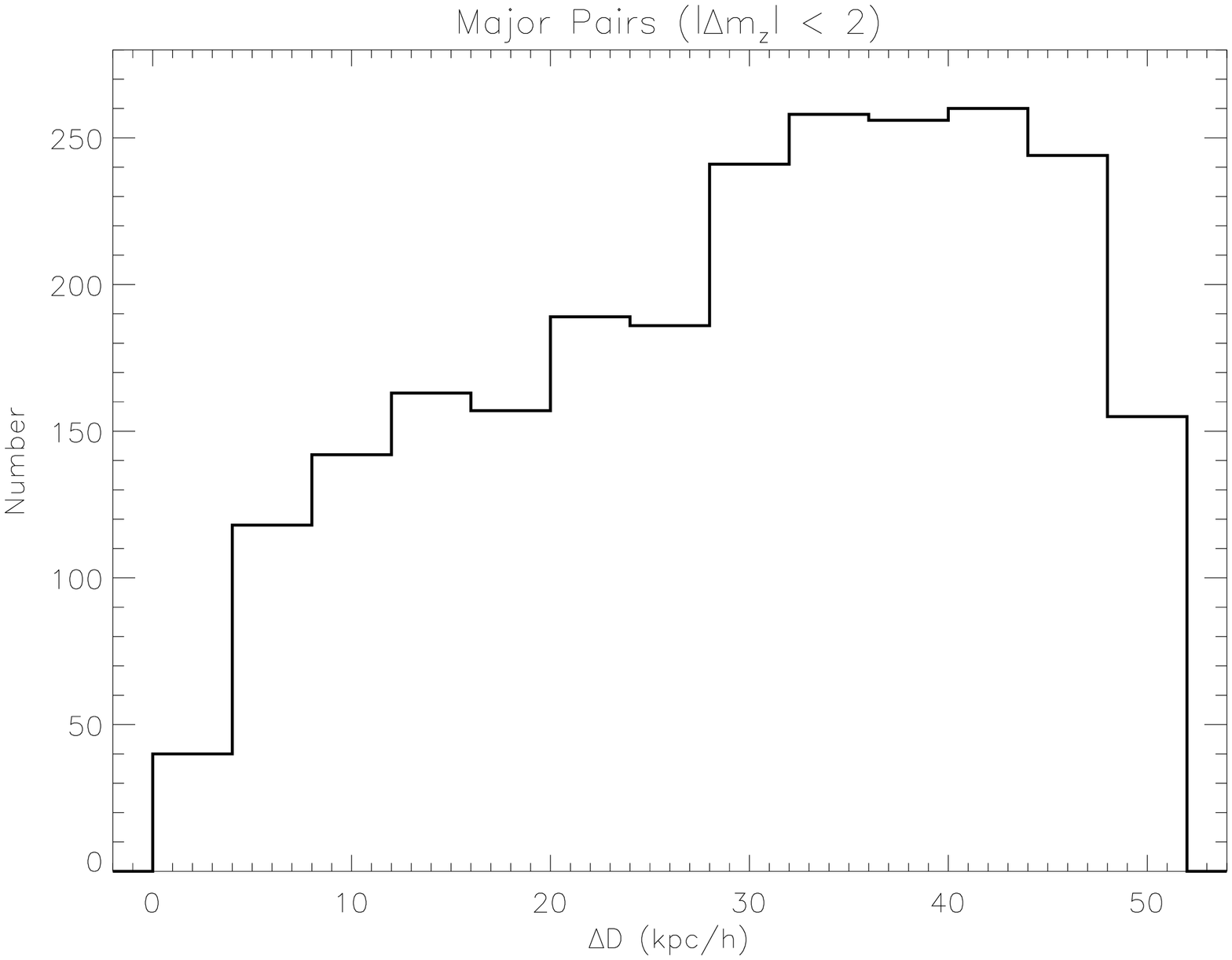}  
\caption{The distribution of projected separations $\Delta D$ for
         the minor pairs (above left) and major pairs (above right).  Both pairs 
	samples are deficient in pairs with small $\Delta D$: the distribution for a 
	complete sample is expected to be flat across all $\Delta D$.} 
  \label{hist-sepp}
\end{figure}

We note that both the major and minor pairs samples are deficient in 
pairs with small projected separations (Figure~\ref{hist-sepp}).  
The true distribution should be nearly flat across $\Delta D$ because 
the slope of the two-point galaxy-galaxy correlation function is $\sim -2$ at 
small separation \citep{padma}.  For the $\Delta D$ distribution of a more
complete sample of galaxy pairs, see \citet{geller06}.

SDSS fiber constraints result in pairs samples that are deficient in 
small projected separations.  This bias is particularly unfortunate for 
studying tidally triggered star formation because the strongest bursts of
 star formation are, of course, expected in pairs with 
small $\Delta D$ \citep[e.g.][]{bgk}.  The results derived from the sample lacking 
small $\Delta D$ pairs are not as strong as would be expected in a sample with 
a flat distribution: there are relatively fewer galaxies with the possibility
for showing the strongest bursts of star formation, and there is a greater
probability for outliers at large $\Delta D$ where the sample is larger.
Re-sampling the distribution to flatten it would in principle provide a better
representation of SSFR at different $\Delta D$, but the reduction in  
sample size would also weaken the statistical significance of the results.
We analyze the SDSS sample without attempting to correct for incompleteness
and take detections of trends as strong indicators of the underlying tidal
triggering.

\subsection{Field Galaxy Sample Selection} \label{fieldsample}

For comparison, we select a field galaxy sample with the same criteria for the 
pairs sample.   The  galaxies lie in the redshift 
range $8,000 < cz < 50,000$~km s$^{-1}$ and have at least 20\% of the total galaxy 
luminosity in the fiber.  We limit the sample  to $b > 40^\circ$ 
to keep the sample size manageable.  We exclude galaxies from dense regions 
($\Delta D < 1.5$~Mpc h$^{-1}$ and $\Delta V < 5000$~km s$^{-1}$ of rich Abell clusters)
from the field sample.  There are 65,570 galaxies 
that meet our selection criteria.  The field galaxy sample properties 
are discussed in \S\ref{properties}.

\section{Sample Properties}  \label{properties}

\begin{figure}[htb!]
\centering
\includegraphics[width=2.3in,angle=90]{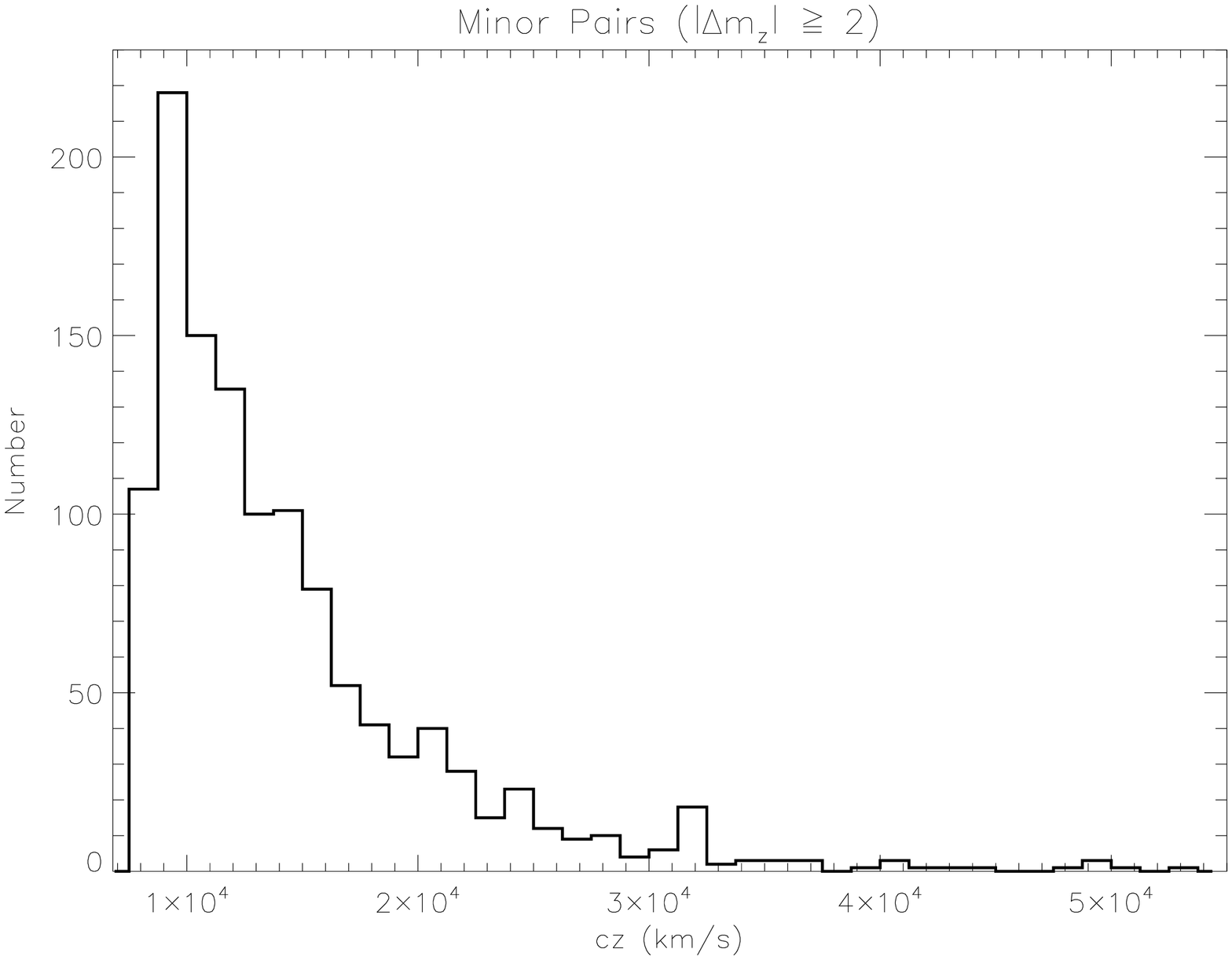}
\includegraphics[width=2.3in,angle=90]{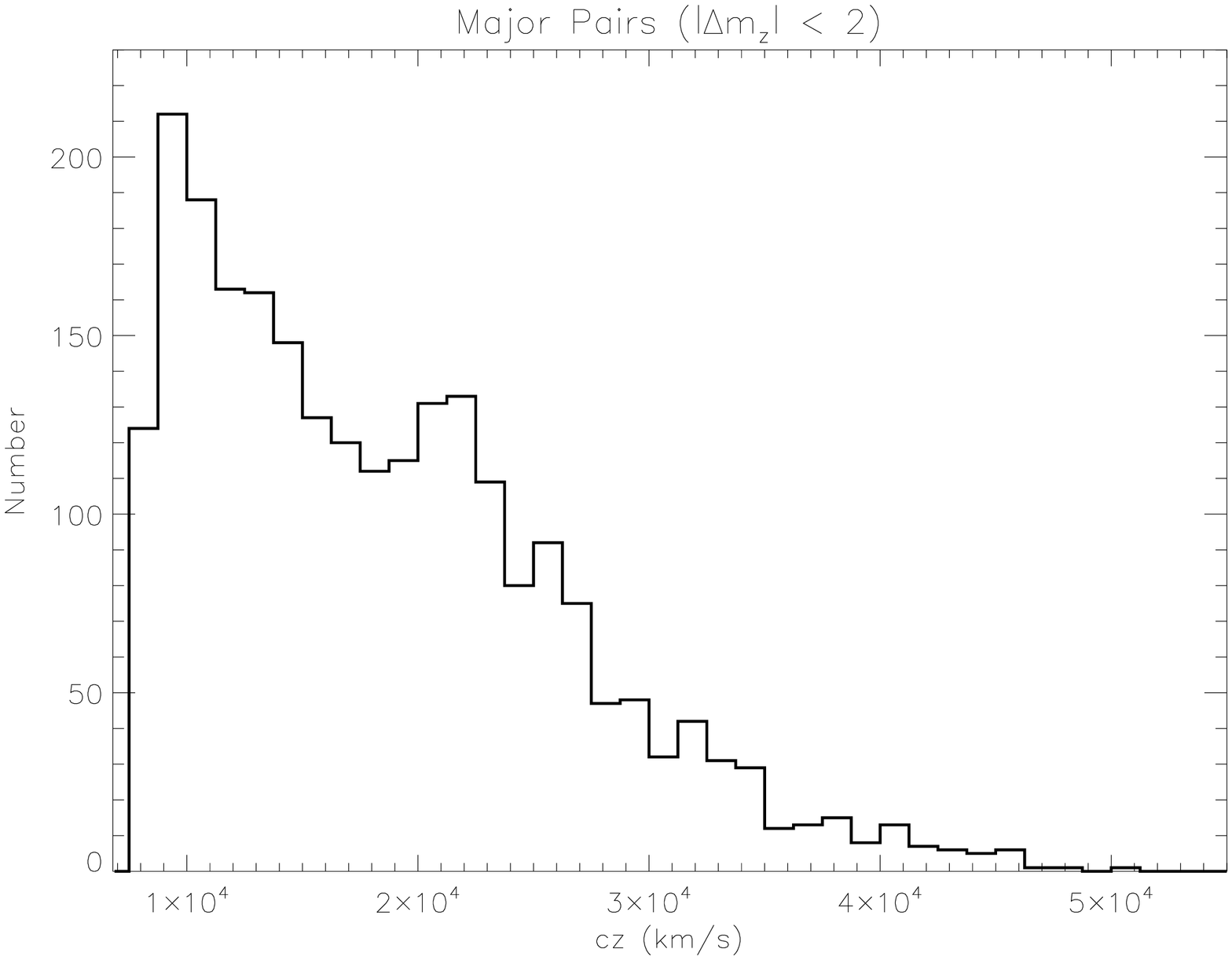}\\
\includegraphics[width=2.3in,angle=90]{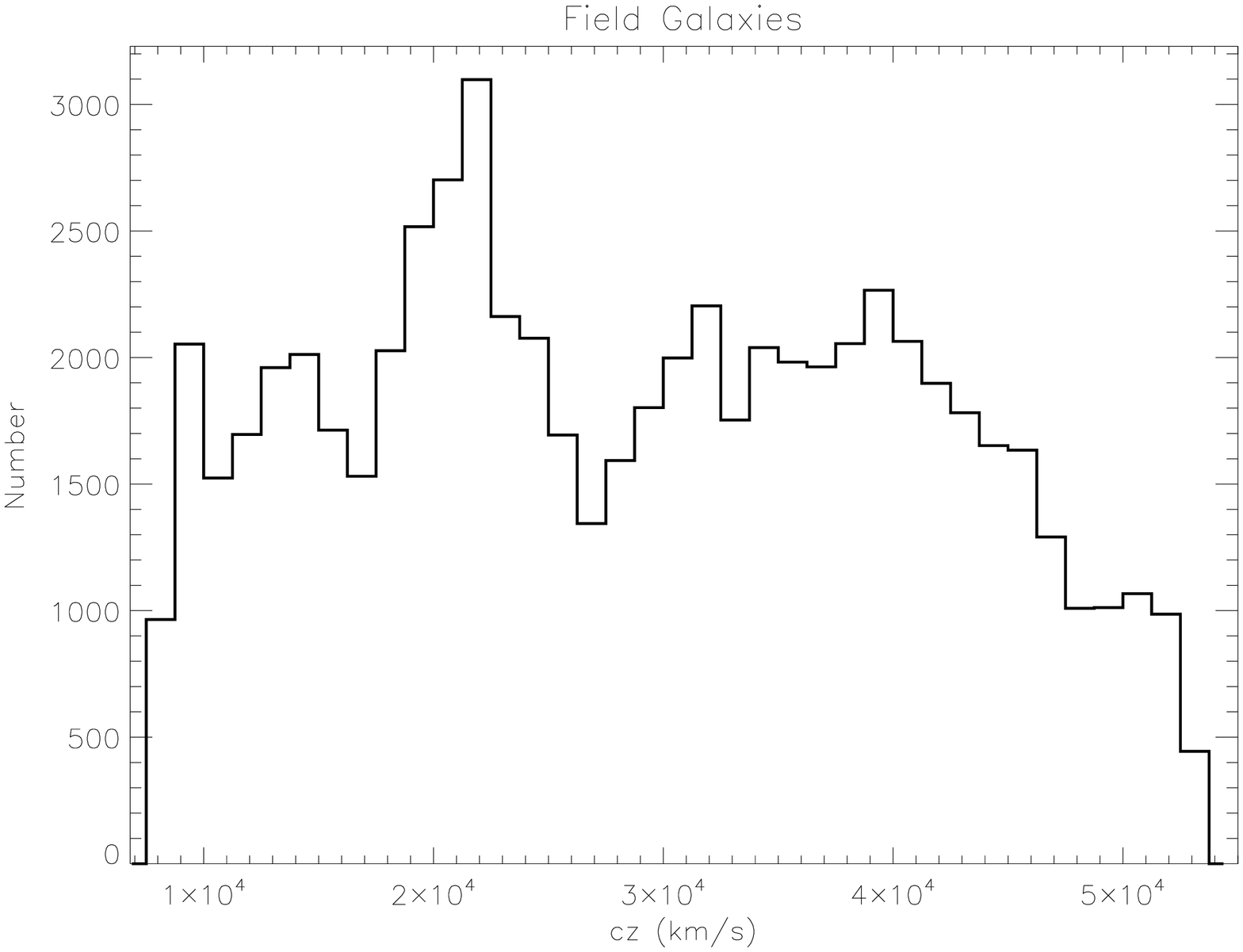}
\caption{The distribution of recessional velocities for the minor
	 pairs (top left), major pairs (top right), and field sample 
	(bottom center).  	The minor pairs must be 
         at lower $cz$ in order for the secondary galaxy to be observable.}
\label{cz-minor}
\end{figure}

The galaxies we select are at low redshift (Figure~\ref{cz-minor}).  
The minor pairs are concentrated in the lowest redshift bins because 
they must be close enough that the faint companion galaxy is brighter than 
the  SDSS spectroscopic magnitude limit, $m_r = 17.77$.

Our sample of major interactions includes pairs with 
$\left | \Delta m_z \right | < 2$, and is weighted toward pairs 
with $\left | \Delta m_z \right | \sim 2$.  This distribution is consistent 
with a greater fractional abundance of relatively low luminosity galaxies.  
The minor pairs sample includes  galaxies within 
$2< \left | \Delta m_z \right | < 6$, and has more galaxies with smaller 
magnitude differences.  The median $\left | \Delta m_z \right |$ for the 
major and minor pairs is 1.5 magnitudes, and the median $\left | \Delta m_z \right |$
of the minor pairs is 2.6.  The galaxies with greater
$\left | \Delta m_z \right |$ are relatively rare because the available
volume for these pairs is small in a magnitude limited sample.

\begin{figure}[htb!]
\begin{center}
\includegraphics[width=2.3in,angle=90]{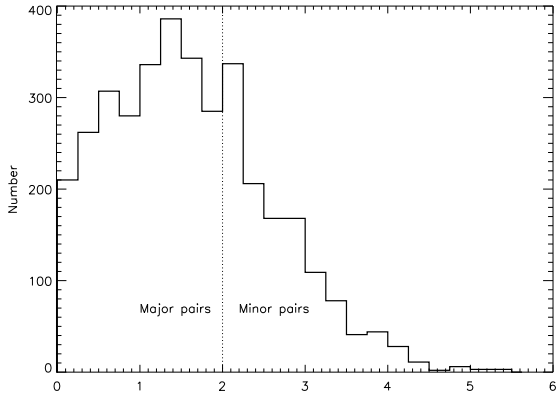}
\caption{The distribution of apparent magnitude differences.  This
	 plot shows both the major ($\left | \Delta m_z \right | <2$) 
	 and the minor ($\left | \Delta m_z \right | \geq 2$) galaxy pairs.
	 The median $\left | \Delta m_z \right |$ for the combined major and 
	minor pairs is 1.5.}
\label{mz}
\end{center}
\end{figure}

\begin{figure}[htb!]
\centering
 \includegraphics[width=2.3in,angle=90]{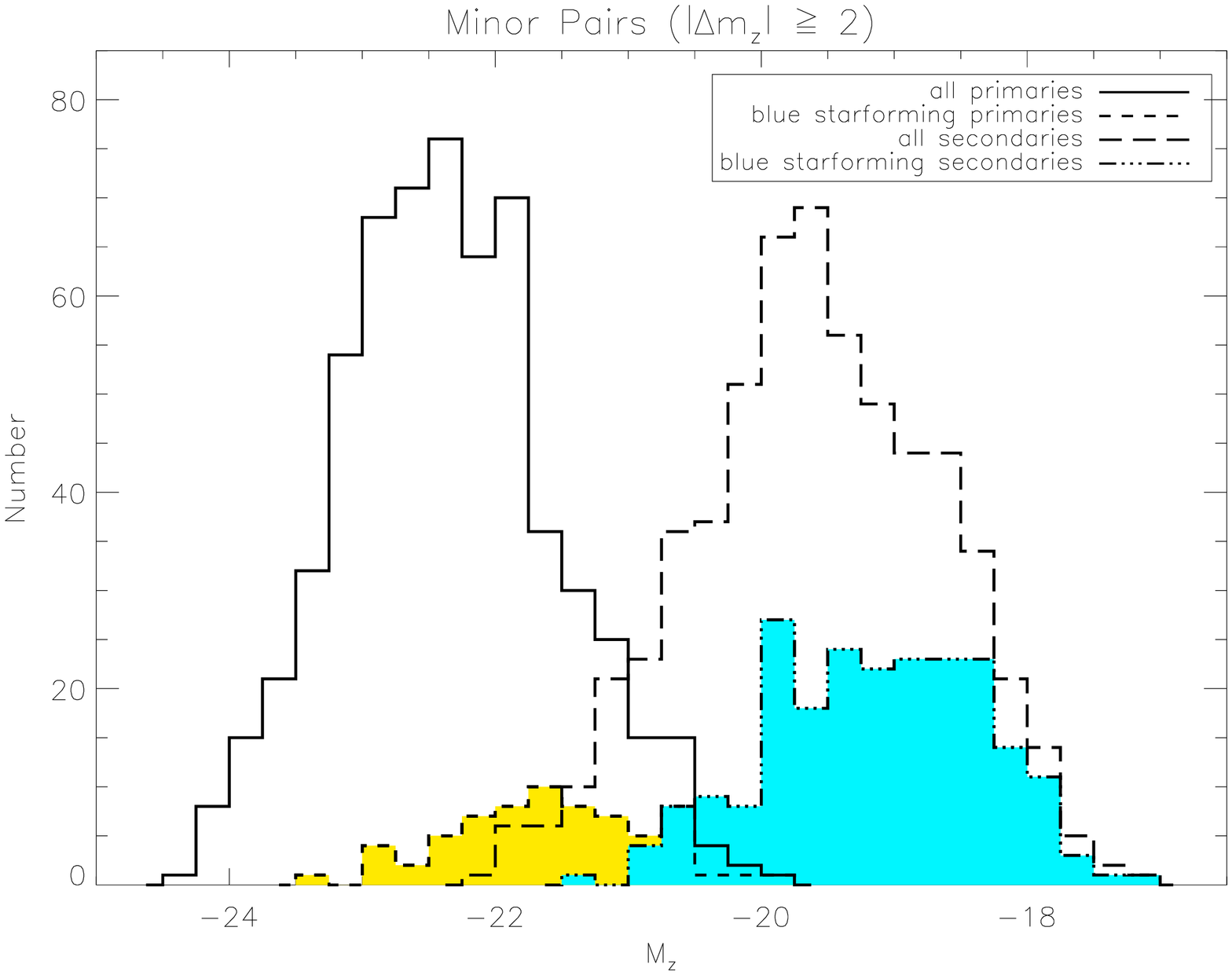}
  \includegraphics[width=2.3in,angle=90]{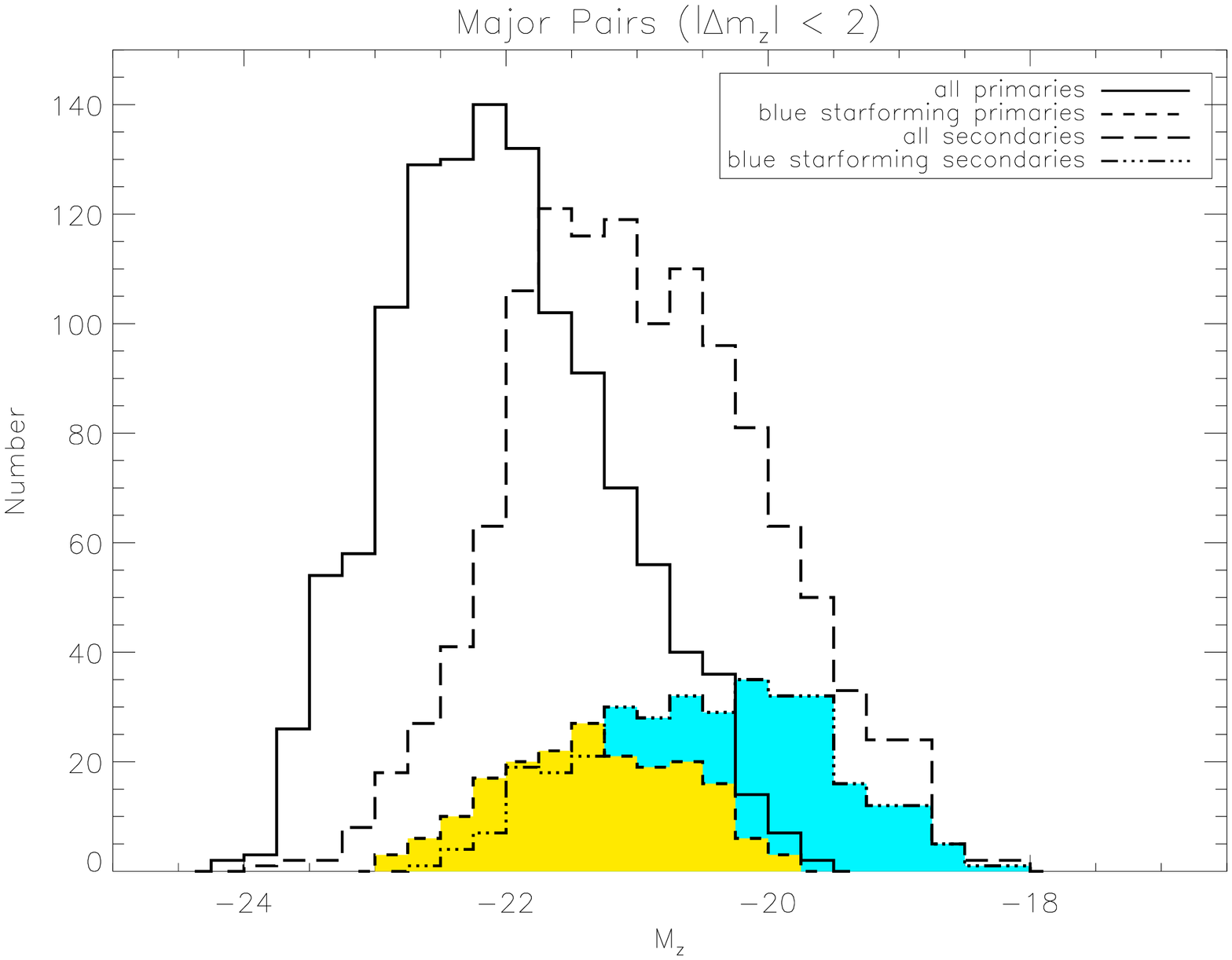}\\
 \includegraphics[width=2.3in,angle=90]{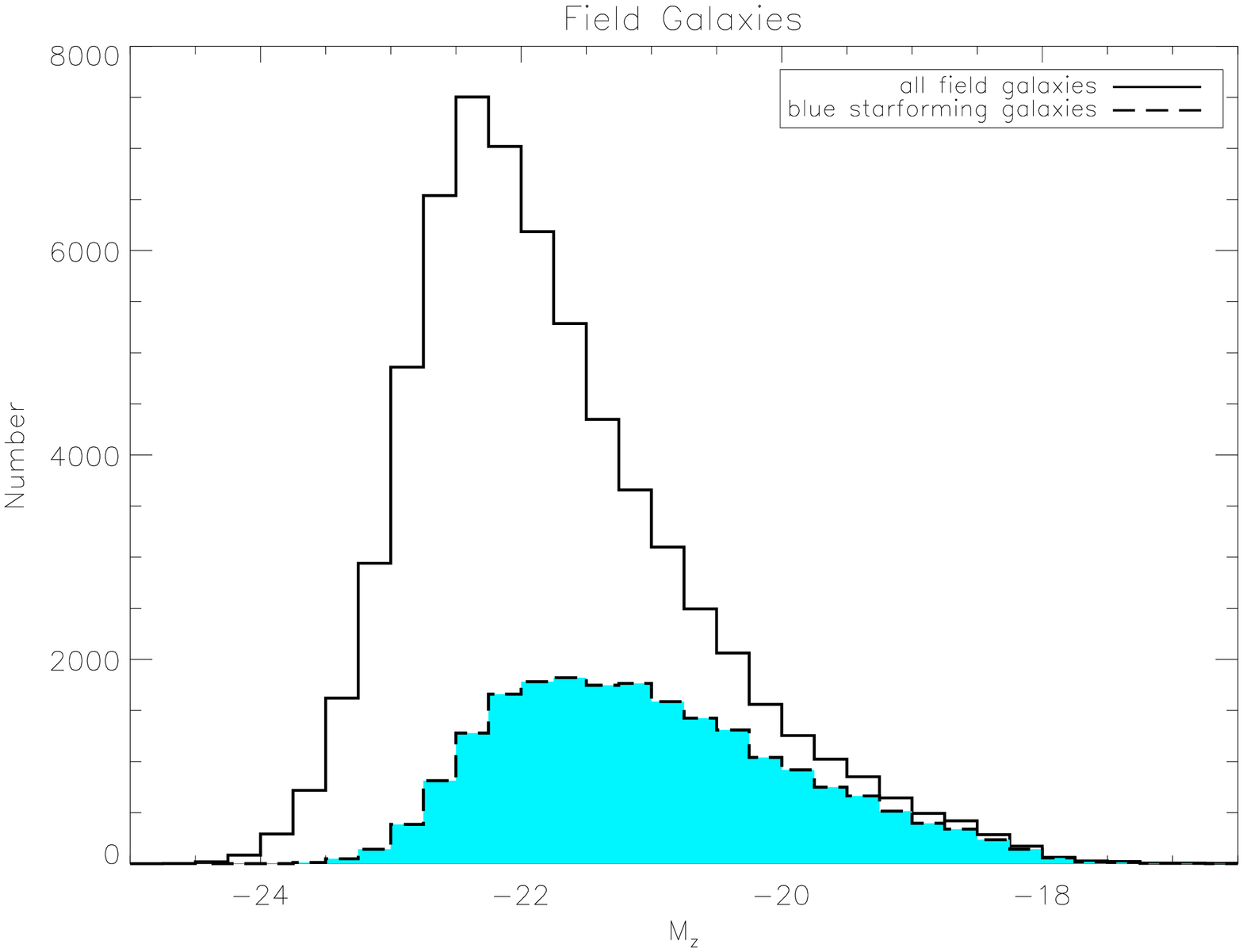}
\caption{Distribution of $M_z$ for the minor pairs (top left), major pairs 
	(top right), and field galaxies (bottom center).}  \label{absz-m}
\end{figure}

Figure~\ref{mz} shows the distribution of magnitude differences 
for both the major and minor pairs samples.
The absolute magnitude distribution of minor pair galaxies is clearly
bimodal; the distribution of $M_z$ of the primary galaxy 
has a median of -22.4, and the distribution for the secondary galaxies
has a median of -19.6 (see Figure~\ref{absz-m}.)  Almost all ($97\%$) of 
the primary galaxies  are brighter than $M_{z}^{*} \simeq -20.7$, and most ($80\%$) 
of the secondaries are fainter [$M_{z}^{*}$ estimated from $M_{r}^{*}= -20.49$; 
\citet{ball06}]. The major pairs sample shows more similar distributions of
$M_z$ for the primary and secondary galaxy, although it is 
still bimodal.  The median $M_z$ of the primary galaxies is -22.1, 
and the median of the secondary galaxies is -21.0.

\begin{figure}[htb!]
\centering
   \includegraphics[width=2.3in,angle=90]{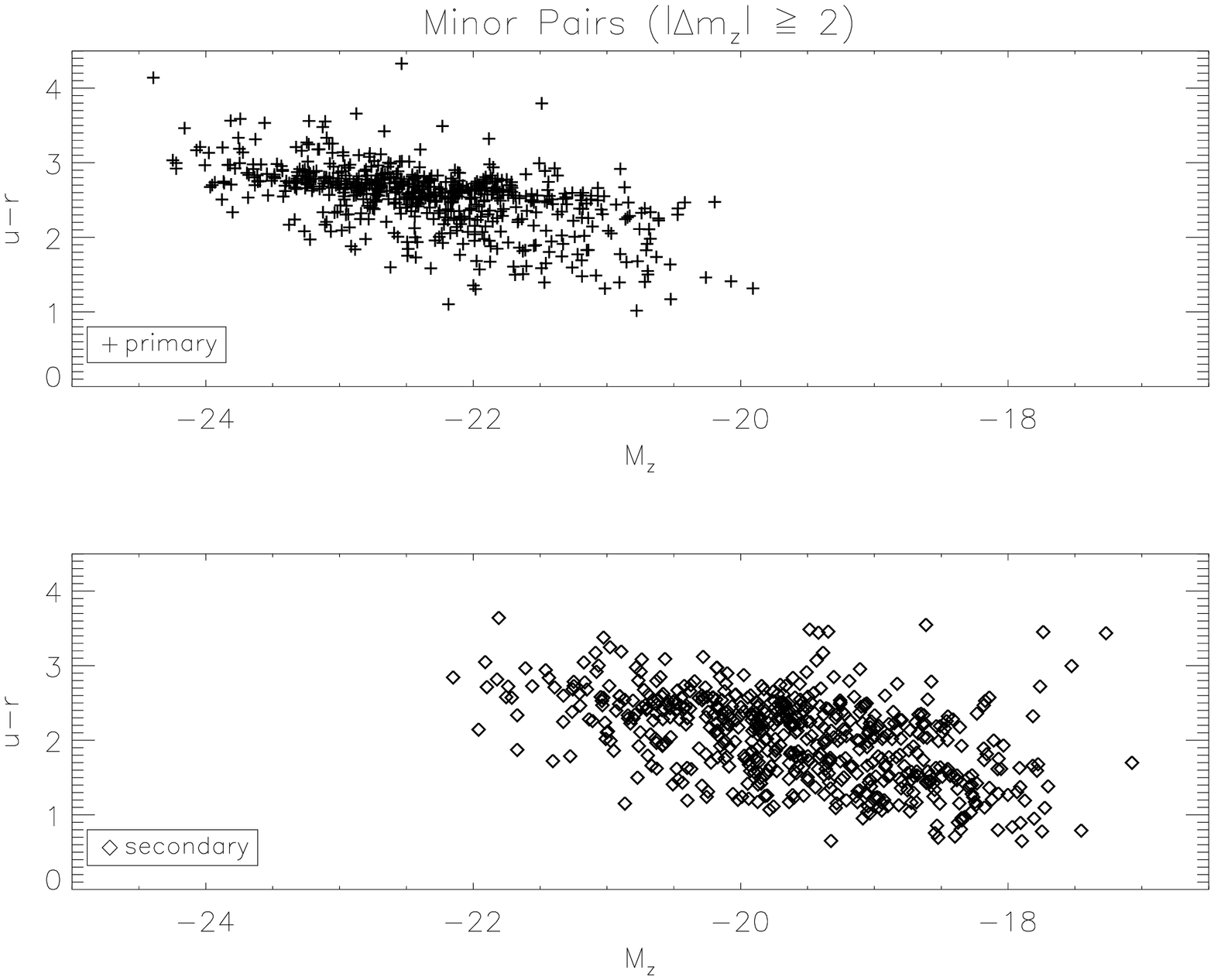}
   \includegraphics[width=2.3in,angle=90]{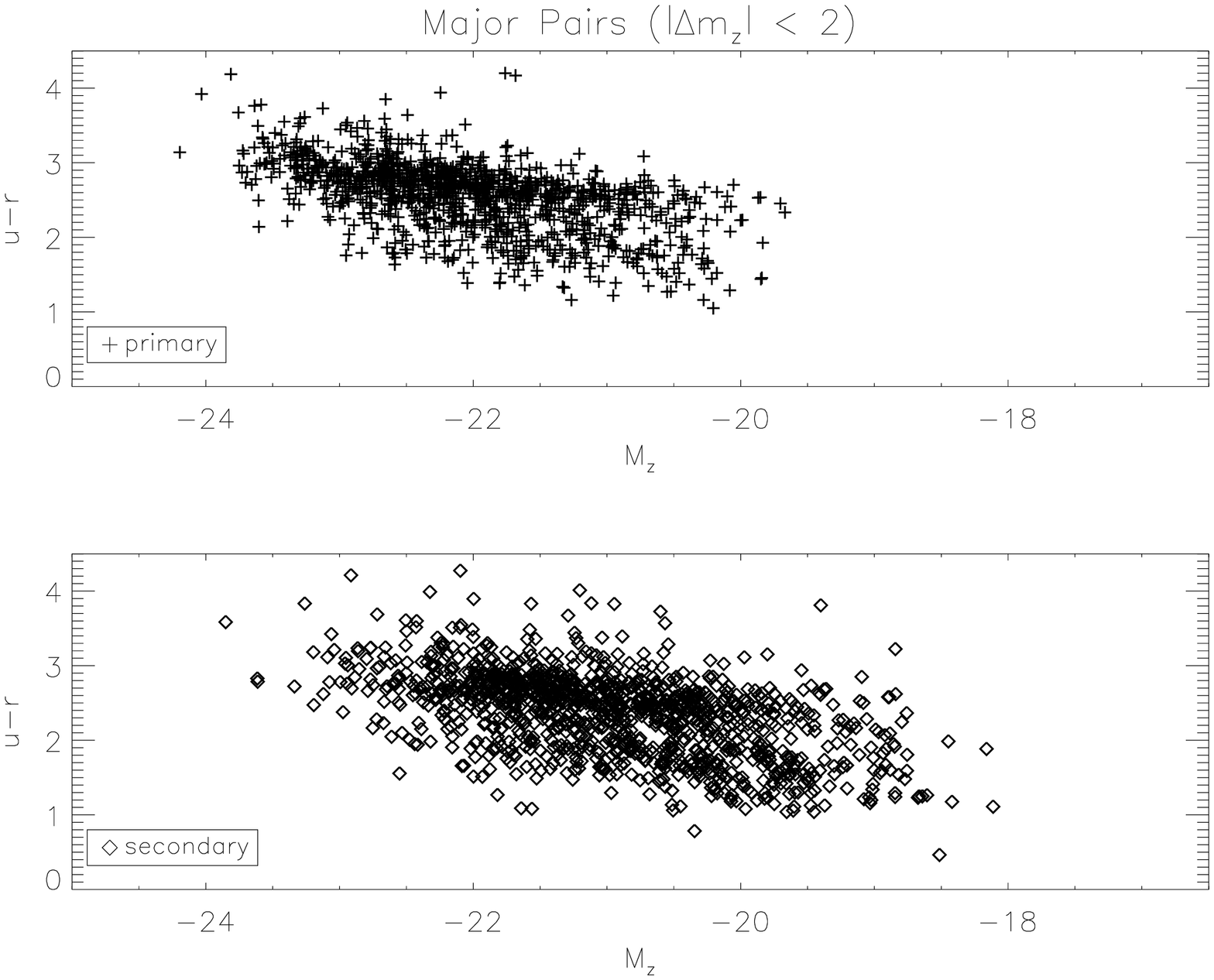}
\caption{The $u-r$ color-magnitude diagram for the galaxies in minor pairs 
	(above left) 	and in major pairs (above right).  }
\label{color-minor}
\end{figure}

\begin{figure}[htb!]
\centering
\includegraphics[width=1.2in,angle=-90]{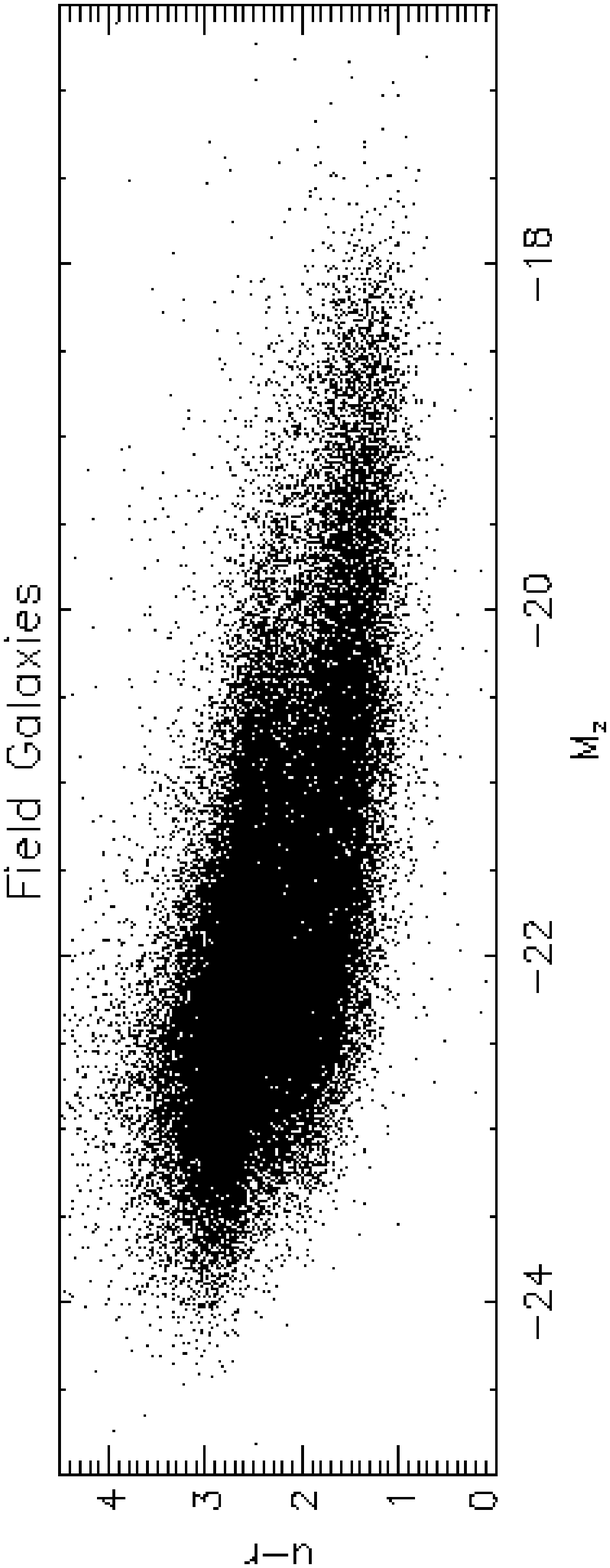}
\caption{The bimodal color distribution can be seen in 
	the field galaxy sample, which spans a wide range of $M_z$.}
\label{color-field}
\end{figure}

We classify our galaxies as red or blue based on the $u-r$ color
and the absolute magnitude $M_r$, as prescribed in \citet{baldry04}.
Because lower luminosity galaxies tend to be bluer, the fraction of blue 
galaxies is related to the mean sample magnitude: 
in our minor pairs sample, $18\%$ (111/608) of the primary galaxies are blue, and
$40\%$ (238/596) of the secondary galaxies are blue.  In the major pairs,
$24\%$ (292/1195) of the primary galaxies are blue and $32\%$ (390/1214) of 
the secondary galaxies are blue.  Figure~\ref{color-minor} shows the 
color-magnitude  diagrams for the minor and major pairs samples;
Figure~\ref{color-field} shows the same diagrams for the field galaxies'.

 Galaxies in our pairs samples with different absolute magnitude 
distributions should have intrinsic differences in star formation rates, 
colors, and AGN fractions.  Our goal is to distinguish the intrinsic
differences from those attributed to the gravitational tidal interaction.

Galaxy colors are well correlated with gas content, star formation rates, 
and other galaxy properties \citep[e.g.][]{sandage,baldry04,bernardi,conselice}. 
Figure~\ref{color-minor} shows that  the secondary galaxies tend to be 
bluer than the primary galaxies.    Because blue galaxies
generally have a greater gas content, they are more likely to show
enhanced central star formation as a result of a gravitational tidal 
interaction.  The simulations of \citet{barnes+hern96} demonstrate the
importance of gas content for tidally triggered star formation.
However, blue galaxies are also more likely to have intrinsically high 
specific star formation rates, independent of the interaction.   Thus we 
look for two measures of the interaction:
(1) enhanced SSFR in pairs overall compared to field galaxies, and
(2) enhanced SSFR as a function of $\Delta D$, which can be attributed
to tidal triggering.

\section{Galaxy Classification: Relative Abundances of \\
Starforming Galaxies, Composites, or AGNs} \label{agns}

\begin{figure}[htb!]
\centering
  \includegraphics[width=2.3in,angle=90]{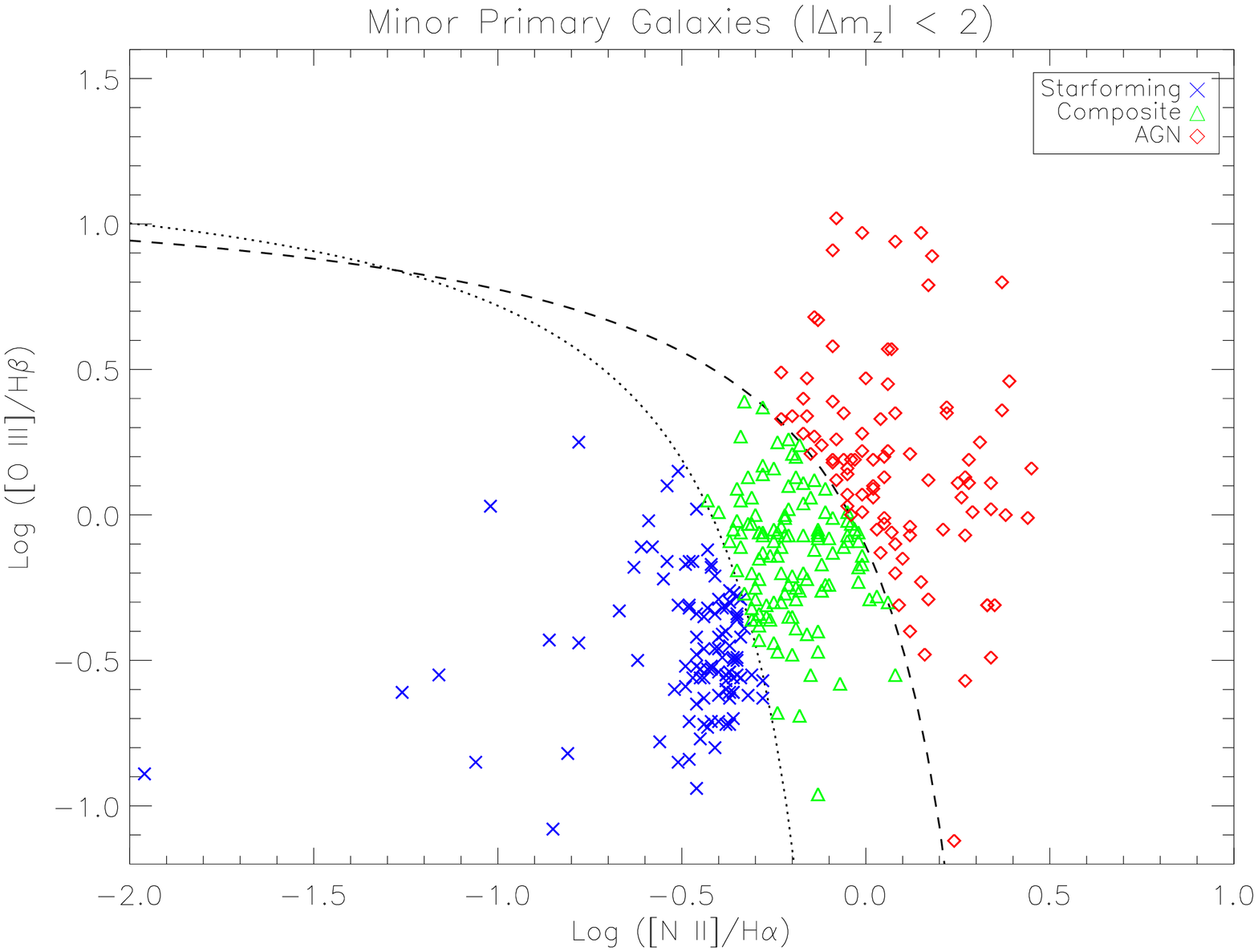}
  \includegraphics[width=2.3in,angle=90]{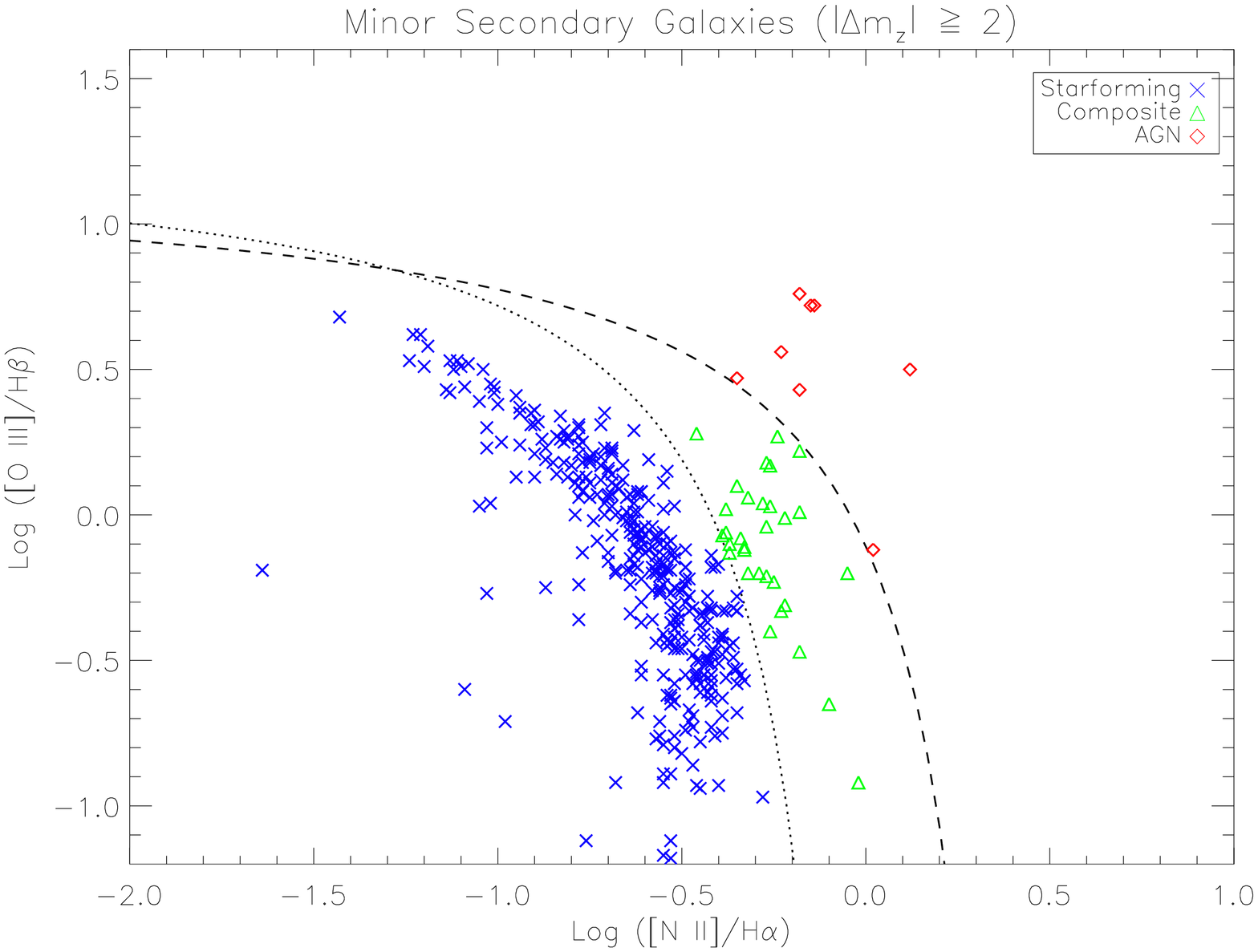}
\caption{Classification of the primary galaxies (above left) and secondary 
	galaxies (above right) 
	in minor pairs as starforming, composite, or  
        AGN using the \citet{kewley06} prescriptions.  The samples
	with greater intrinsic luminosities have larger AGN fractions
	(Table~\ref{frac-agn}). }
\label{class-minor}
\end{figure}

\begin{figure}[htb!]
\centering
  \includegraphics[width=2.3in,angle=90]{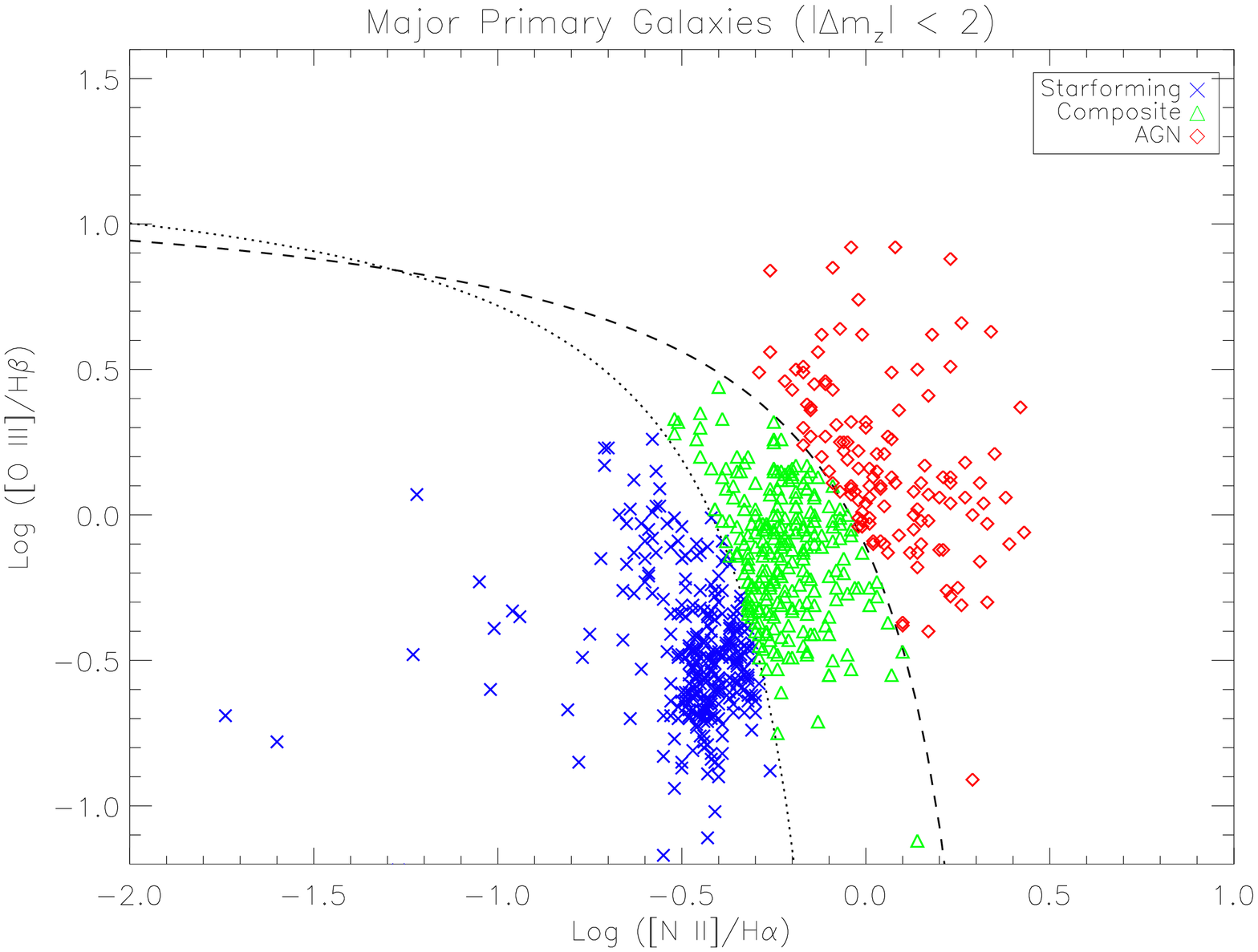}
  \includegraphics[width=2.3in,angle=90]{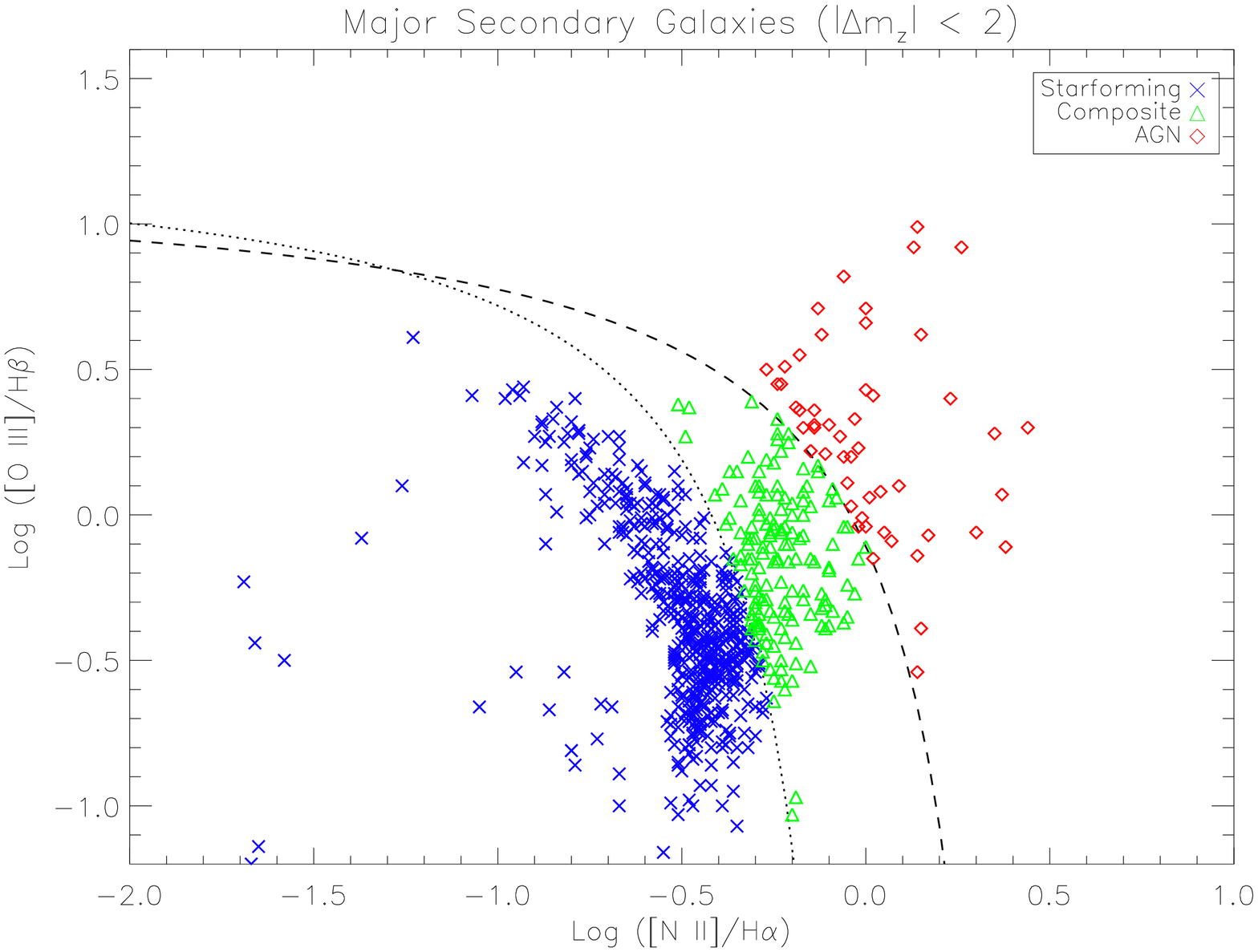}
\caption{Classification of the  primary galaxies (above left) and secondary 
	galaxies (above right) in major pairs as starforming, composite, or  
        AGN using the \citet{kewley06} prescriptions.  The samples
	with greater intrinsic luminosities have larger AGN fractions
	(Table~\ref{frac-agn}). }
\label{class-major}
\end{figure}

\begin{figure}[htb!]
\centering
 \includegraphics[width=2.3in,angle=90]{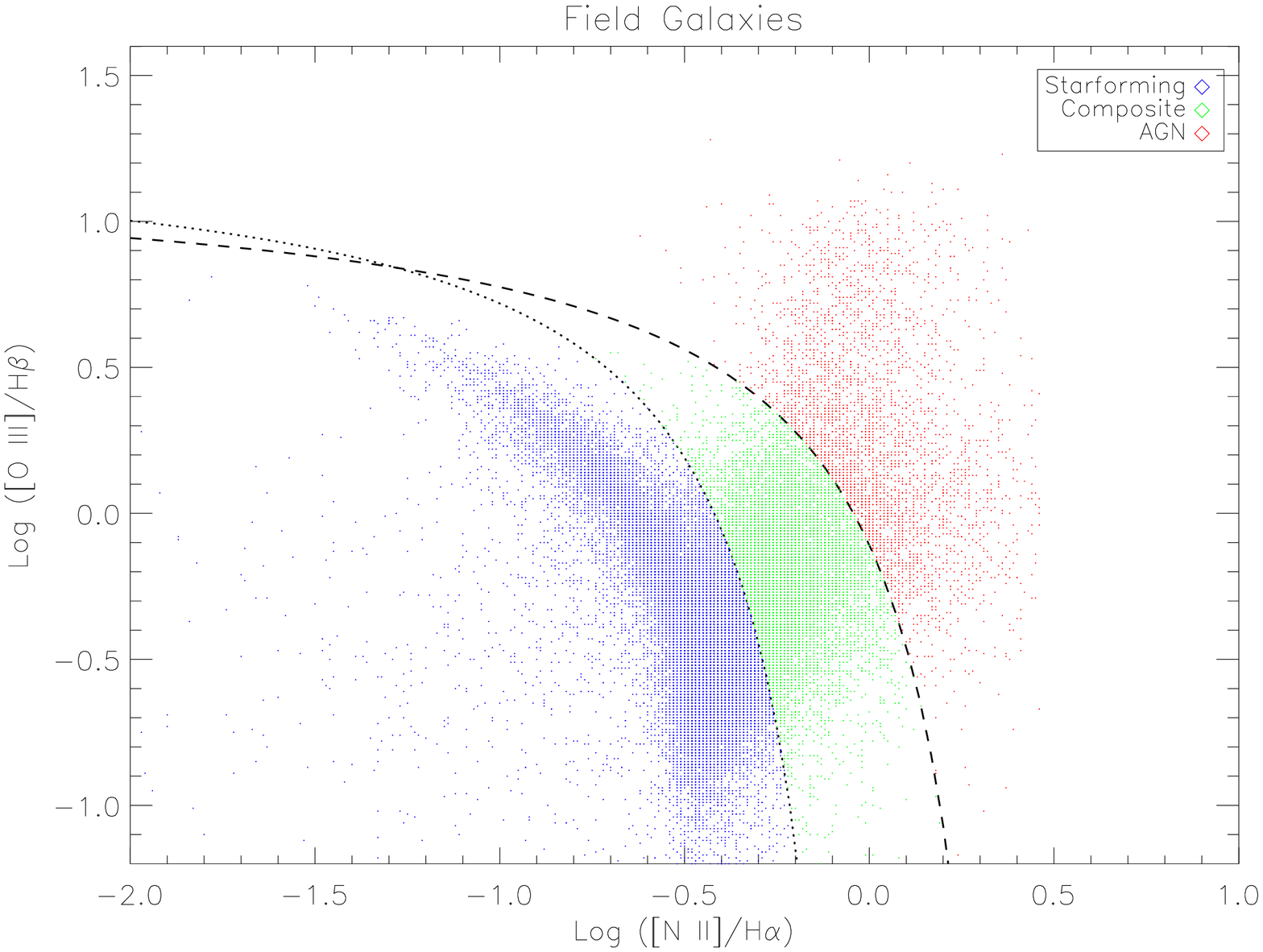}
\caption{Classification of the field galaxies as starforming, composite, or  
        AGN using the \citet{kewley06} prescriptions.}
\label{class-field}
\end{figure}

We use emission line ratios to  classify  galaxies as starforming galaxies, composites, 
or AGNs (Figures~\ref{class-minor},\ref{class-major}, and \ref{class-field}), 
according to the definitions in \citet{kewley06}.  Separating the starforming 
galaxies from the AGNs is necessary for a clean measurement of galaxy
star formation rates.  Our SSFR indicator is a function
of the H$\alpha$ emission, and thus we select for galaxies in which
starforming regions, not AGN, dominate the H$\alpha$ emission.
We maximize our sample size of starforming galaxies by accepting
cases that potentially harbor a weak AGN, where the dominant 
contribution to the H$\alpha$ emission probably originates from 
starforming regions.  To that end, we require a signal-to-noise
ratio $> 3$ in H$\alpha$ and H$\beta$ to ensure an accurate
reddening correction to the H$\alpha$ star formation rate.  We do 
not impose limits on the signal-to-noise in \othree\ and \ntwo\ because
weak \othree\ and \ntwo\ emission lines indicate weak or no AGN activity.
The scatter in small \othree/H$\beta$ and small \ntwo/H$\alpha$ is due to
noisy values of small \othree\ and \ntwo.

Our aperture requirement of at least $20\%$ of the galaxy light landing
in the fiber introduces a slight bias toward centrally
concentrated galaxies.  Centrally concentrated galaxies are 
more likely to be early-type \citep{strateva01}, and to harbor an 
AGN \citep{alonso07}.  However, because the spectra include a significant 
fraction of the galaxy light, it is harder to observe weak AGNs against 
the background galaxy.  Because we cannot eliminate these selection
effects, we control for them by maintaining the same selection criteria
for the minor pairs, major pairs, and field samples.

The different pairs samples show very different AGN fractions, ranging
from $2.1\%$ in the minor secondary galaxies to $29\%$ in the minor
primary galaxies.  The field galaxies are $8\%$ AGNs.
The subsets containing the more luminous galaxies 
have a larger AGN fraction; the median $M_z$ of all primary galaxies 
in the minor pairs sample is -22.4, compared to a median of 
$M_z = -19.6$ for all secondary galaxies in the minor pairs sample 
(Figure~\ref{absz-m}).  Our AGN fractions for the various 
pairs samples are consistent with the observed trend of increasing 
AGN fraction with increasing galaxy luminosity \citep{kauffmann03b,best05}.

We compare the AGN fraction of the galaxies in our pairs samples 
with matched subsets of galaxies from our field sample.  We select a 
subset of field galaxies having
the same distribution of $M_z$ and the same number of red and blue
galaxies as in the corresponding pairs sample.  Repeating the 
random selection for 5,000 trials, we find the average percentages of
starforming galaxies, composites, and AGNs in our representative sets of field
galaxies.  Table~\ref{frac-agn} shows the classification for the pairs 
samples  and for the matched sets of field galaxies.

\begin{table}[htb]
\caption{Galaxy classification in minor pairs, major pairs, and field galaxies.
\label{frac-agn} }
\begin{tabular}{lllllll}
\tableline 
\tableline
Sample & Sample Size & Classified\tablenotemark{a} & Starforming & Composite & AGN & AGN in Field\tablenotemark{b} \\
\tableline
Minor primary & 608 & 324 & $34\%$ & $38\%$ & $29\%$ & $14\%$ \\
Minor secondary & 596 & 377 & $90\%$ & $8.2\%$ & $2.1\%$ & $2.0\%$ \\
Major primary & 1195 & 686 & $45\%$ & $34\%$ & $18\%$ & $12\%$ \\
Major secondary & 1214 & 709 & $71\%$ & $22\%$ & $7\%$ & $6\%$ \\
Total field & 65,570 & 41,894 & $64\%$ & $27\%$ & $8\%$ & ... \\
\tableline
\end{tabular}
\tablenotetext{a}{Classified galaxies must have measurable emission in [N II], 
                  H$\alpha$, [O III], and H$\beta$, and the H$\alpha$ and H$\beta$
                  flux have signal-to-noise ratios $>3$.}
\tablenotetext{b}{The matched field sample represents the average over 5,000
                  random draws of field galaxies with the same distribution
                  of $M_z$ and the same numbers of red and blue galaxies that
                  are in the corresponding pairs sample.}
\end{table}

The primary galaxies in both the major pairs and the minor pairs show 
statistically significant increases in AGN fraction (a $5\sigma$ increase 
for the minor primaries and a $7\sigma$ increase for the major primaries.)  
Because the field galaxy subsets are 
selected to have  the same distribution of $M_z$ and the same number of 
red and blue galaxies as their counterpart pairs samples, the correlation 
between galaxy luminosity and AGN fraction cannot explain the differences.  
 For the primary galaxies in the minor pairs, it may be the case 
that more than one secondary galaxy is present.  We do not have the data to
count the number of minor companions.  However, our selection does
require that the primary galaxies have no major companions within 
$\Delta D = 50$~kpc h$^{-1}$ and $\Delta V = 500$~km s$^{-1}$ (\S\ref{selection}).
It is clear that having at least one minor companion increases the odds
that a galaxy harbors an AGN.

The increased AGN fraction in our pairs samples is generally consistent with 
the recent work of \citet{alonso07}, who find a $10\%$ excess of AGN in 
SDSS pair galaxies compared to their reference sample.  However, our 
different sample selection prohibits a direct comparison with their work
(the Alsono et al. sample includes only pairs that exhibit signs of tidal 
interaction on visual inspection.)   Numerous prior works have investigated 
the association of AGN activity and galaxy interactions 
\citep[e.g. ][]{dahari85,keel85,fuentes88,kelm}, and there is some debate 
over the role of galaxy interactions in triggering AGN activity.  Recent 
observational studies by \citet{hennawi06} and \citet{serber06} suggest 
that AGN activity is associated with local companions, in agreement with 
the numerical modeling of \citet{kauffman+hae00,hopkins06}.

\begin{figure}[htb!]
\begin{center}
\includegraphics*[width=1.35in,angle=270]{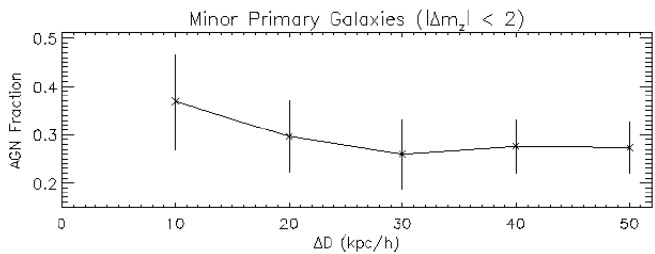}
\caption{The fraction of galaxies classified as AGN according to the
	 definitions in \citet{kewley06} for the primary galaxies
	 in minor galaxy pairs.}
\label{agn-sep}
\end{center}
\end{figure}

A plot of the AGN fraction as a function of
$\Delta D$ also hints at enhanced AGN activity for the primary galaxies in
the minor pairs (Figure~\ref{agn-sep}).  We detect the trend in AGN 
fraction versus $\Delta D$ at only the 1 to 2$\sigma$ level.  
A larger data set of pairs at
small projected separation would help to determine the relationship
between AGN activity and gravitational tidal interactions.  We note
that the size of the spectroscopic aperture influences our ability to
detect AGNs against the more luminous host galaxies; our sample is 
selected to have galaxies with $\geq 20\%$ of the light in the fiber.
Nuclear spectra would allow for the classification of weaker AGNs.  
Considering our limited ability to identify AGNs, and the 
dearth of pairs at small $\Delta D$, the observed correlation of 
AGN fraction with $\Delta D$ is potentially very interesting, but 
requires further investigation with a larger data set.

The galaxies in major pairs do not show an obvious trend in AGN fraction
versus $\Delta D$.  The effects of our aperture requirement and 
the absence of pairs at small $\Delta D$ may be washing out the results
for the major pairs, although it is unclear why the trend for the minor 
primaries can be measured despite these effects.  We note that a smaller 
percentage of galaxies in the major pairs are classified as AGN across all 
$\Delta D$ compared to  the minor pairs (Table~\ref{frac-agn}), perhaps 
due to their different distributions of intrinsic luminosities.  A 
larger sample of major pairs at small $\Delta D$ would help to clarify
the relationship between AGN fraction and pair properties.

\section{Distribution of Specific Star Formation Rates in Pairs 
and Field Galaxies} \label{sfr}

\begin{figure}[htb!]
\begin{center}
\includegraphics[width=2.3in,angle=90]{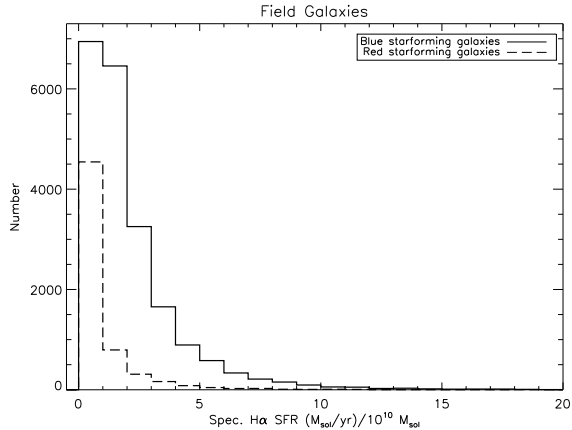}
\caption{Distribution of SSFR for the entire field galaxy sample.
         There are a few objects with 20 $<$ SSFR $>$ 100, which
         make up $0.20\%$ (42/21490) of the blue starforming galaxies,
         and $5.5\times10^{-2}\%$ (3/5411) of the red starforming galaxies.}
\label{field-histsf}
\end{center}
\end{figure}

After removing the AGN and composite objects from our pairs and field samples,
we investigate the star formation properties for the starforming galaxies.
References to SSFR in our samples from hereon refer to the starforming galaxies
only, unless specified otherwise.  We use the \citet{hopkins03} prescription 
for calculating the H$\alpha$ star formation rate based on the SDSS observed 
H$\alpha$ flux.  The H$\alpha$ SFR of Hopkins et al. utilizes
the H$\alpha$ luminosity ($L_{H\alpha}) $calibration from \citet{kennicutt98}:
\begin{equation} 
SFR_{H\alpha}(M_{\sun} \mbox{ yr}^{-1}) = \frac{L_{H\alpha}}{1.27\times10^{34} W}.
\end{equation}
To calculate the $L_{H\alpha}$ from the SDSS H$\alpha$ flux, Hopkins et al.
prescribe corrections for aperture size and for obscuration by dust in the target
galaxy.  We use the Balmer ratio to correct for dust obscuration and the
``alternative aperture correction'' term that is proportional to 
$r_{Petro} - r_{fiber}$ (see Equation~B3 in Hopkins et al.)
To derive the specific star formation rate (SSFR) from the SFR, 
we normalize the SFR by the $M_z$ of the galaxy, where the normalization 
factor assumes $1 M_{\sun} = 1 L_{\sun}$ in the $z$-band and
$M_{z(\sun)} = 4.52$ \citep{yasuda01}.  Figure~\ref{field-histsf} shows
the distribution of SSFR in the field galaxies sample.  

As shown in \S\ref{properties}, the major and minor pairs, and the primary 
and secondary galaxies within those pairs have different characteristic colors 
and absolute magnitudes.  To make a fair comparison of SSFR distributions 
between the various pairs samples and the field sample, we select subsets of 
the field sample with the same distributions of absolute magnitudes and the same
proportion of red and blue galaxies as the relevant pairs sample.  We repeat
the selection of the matched field galaxy subset for 5,000 random
trials, and then measure the Kolmogorov-Smirnov (K-S) probability that the 
distributions of SSFR are drawn from the same parent distribution,
averaged over the 5,000 random trials.

\begin{figure}[htb!]
\centering
   \includegraphics[width=2.3in,angle=90]{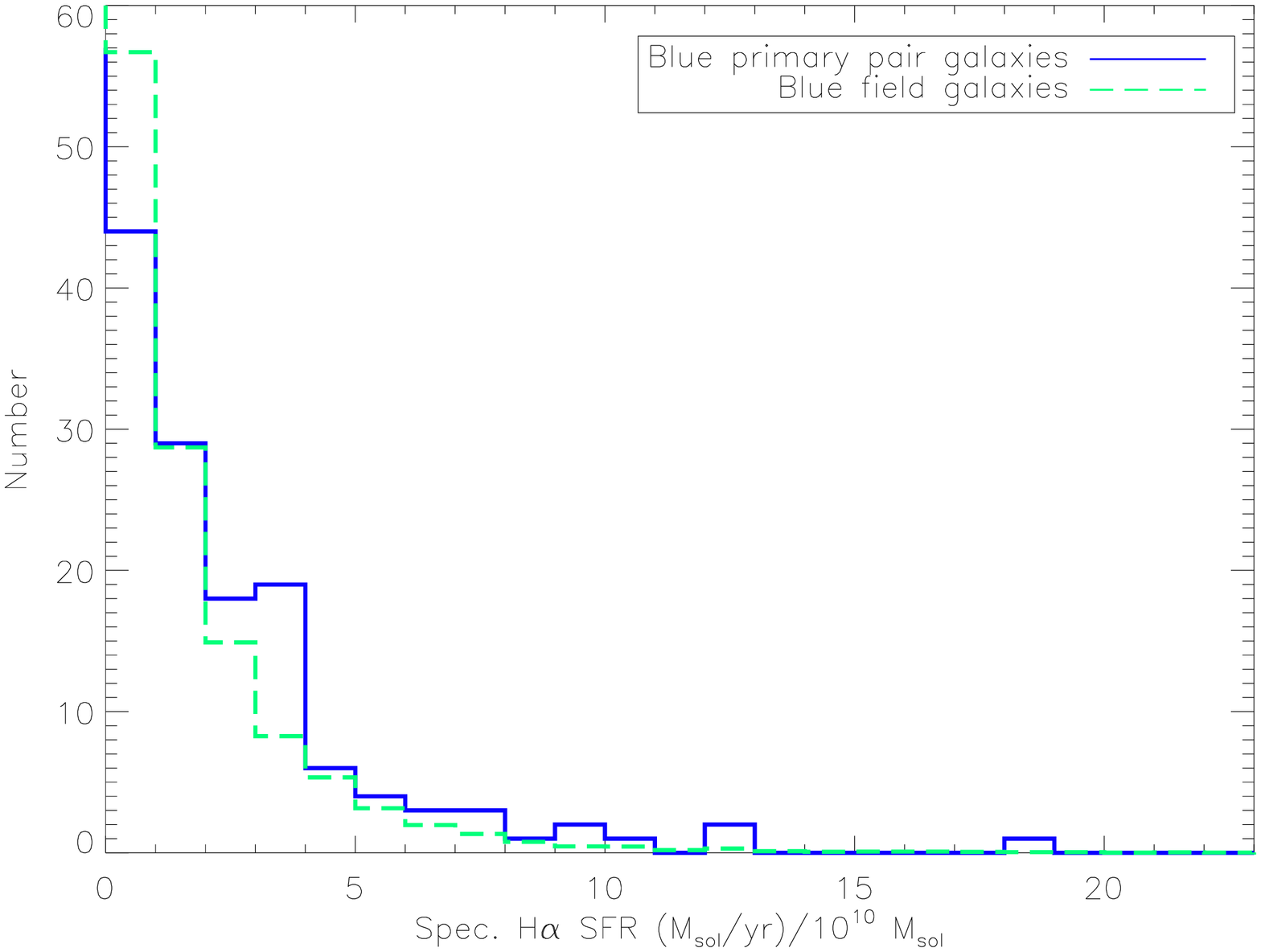}
   \includegraphics[width=2.3in,angle=90]{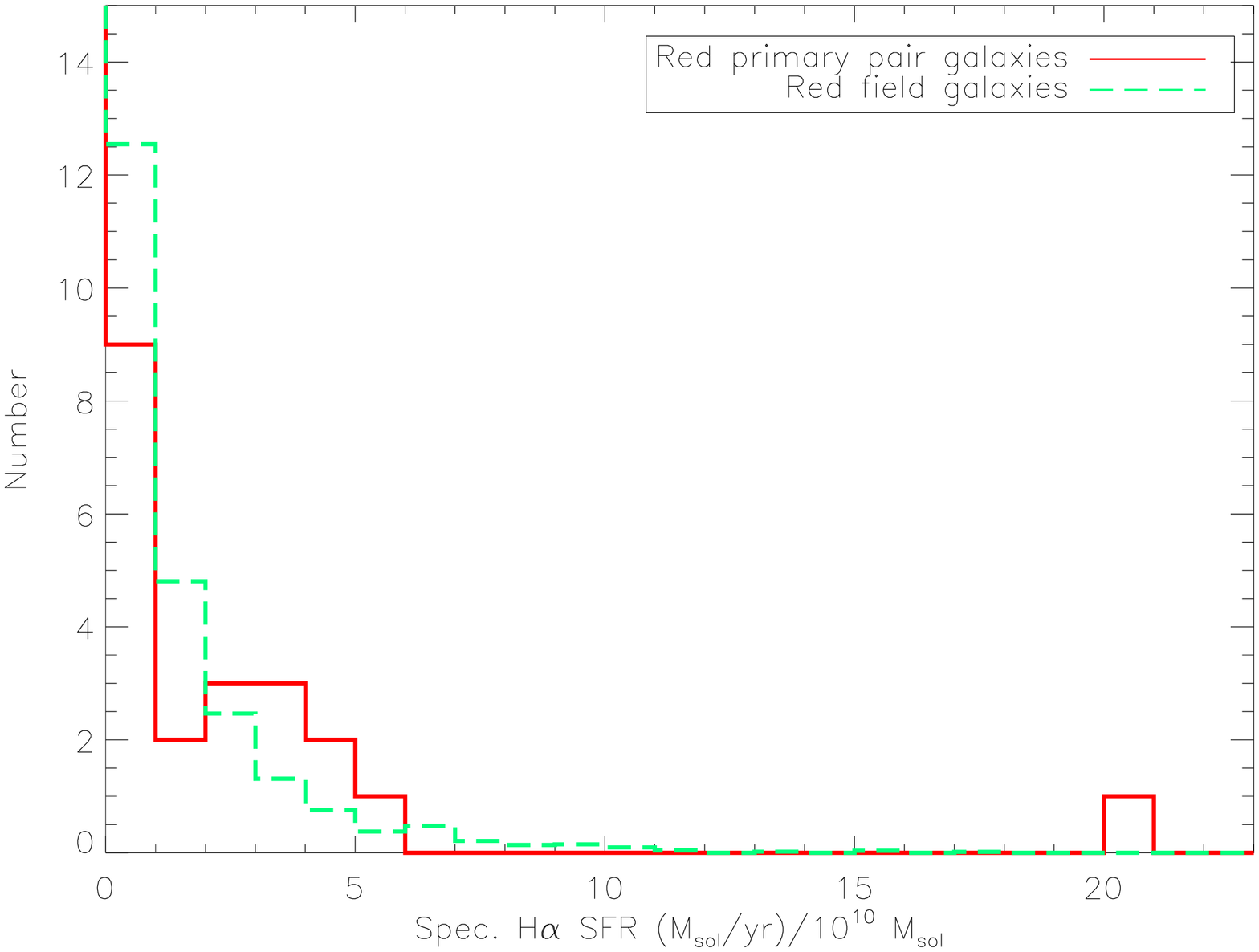}\\
\caption{Distribution of SSFR for the primary galaxies in the
	 major pairs sample that are blue (above left) or red (above right), 
	along with the average distribution of
	 SSFR for the matched sets of field galaxies.}
\label{pbj}
\end{figure}

\begin{figure}[htb!]
\centering
  \includegraphics[width=2.3in,angle=90]{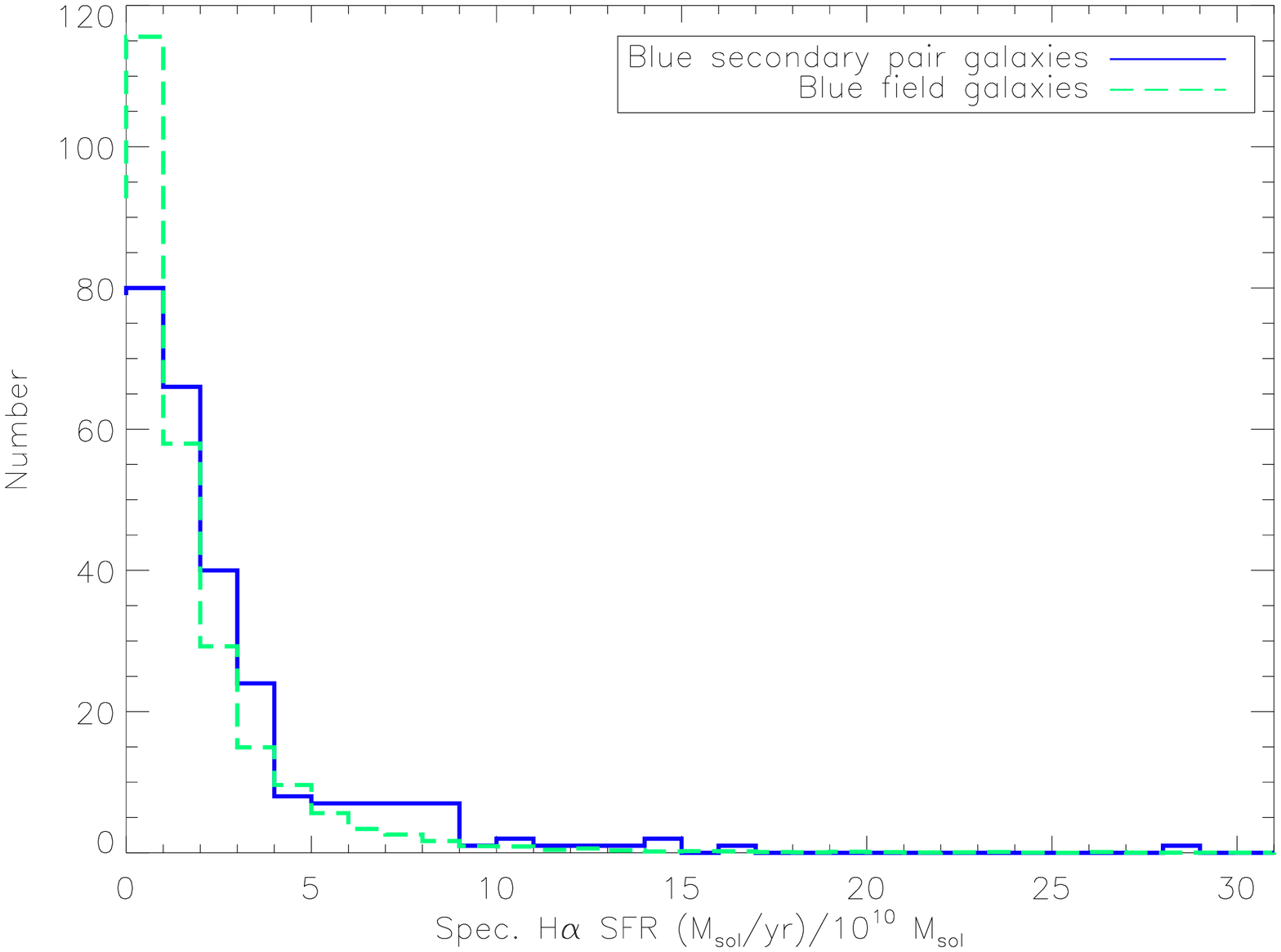}
   \includegraphics[width=2.3in,angle=90]{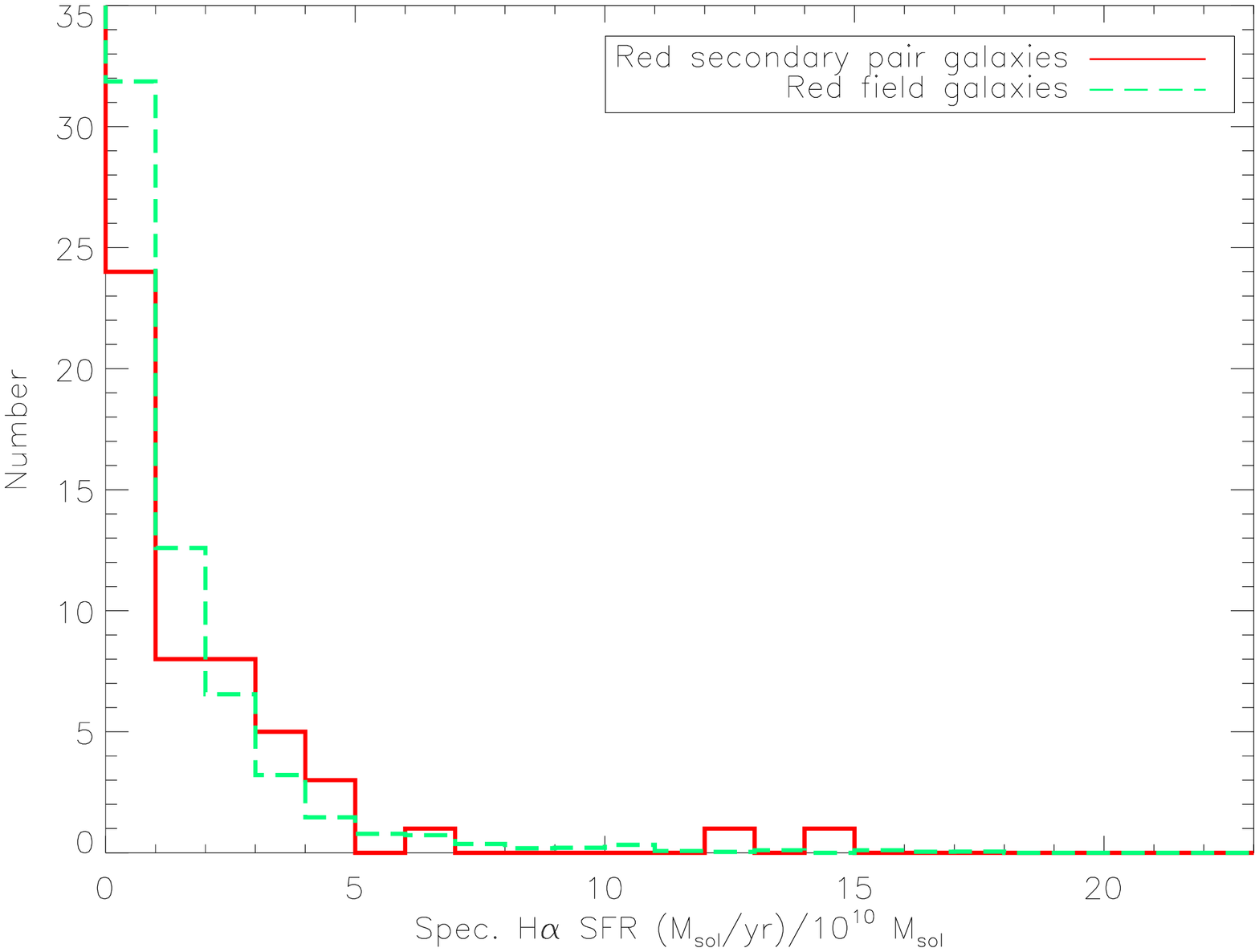}
\caption{Distribution of SSFR for the blue secondary galaxies (above left)
	and the red secondary galaxies (above right) in the
	 major pairs sample, along with the average distribution of
	 SSFR for the matched sets of field galaxies.}
\label{sbj}
\end{figure}

Both primary and secondary galaxies in the major pairs sample have  
distributions of SSFR that differ from the matched sets of field galaxies.
For the blue starforming primary galaxies, the K-S probability of 
the SSFR deriving from the same parent sample as  the matched
subset of field galaxies is $P_{KS} = 5.5 \times 10^{-2}$; for the 
secondary blue starforming galaxies, the K-S probability is 
$P_{KS} = 7.0 \times 10^{-3}$.
As a sanity check, we measure the K-S probability of the magnitude
distributions deriving from the same parent samples, and find 
$P_{KS} = 0.97$ and 0.83 for the primary and secondary blue
starforming galaxies, confirming that the magnitude distributions 
are consistent with having the same parent distribution.
The blue starforming galaxies, both primary and secondary,
 tend to have a larger proportion of high SSFRs, compared to the 
field galaxies.  Figures~\ref{pbj} and \ref{sbj} show the SSFR distributions 
for the red and blue major pair galaxies, along with the average SSFR distributions 
for the 5,000 random trials of matching field galaxy subsets.

\begin{figure}[htb!]
\centering
  \includegraphics[width=2.3in,angle=90]{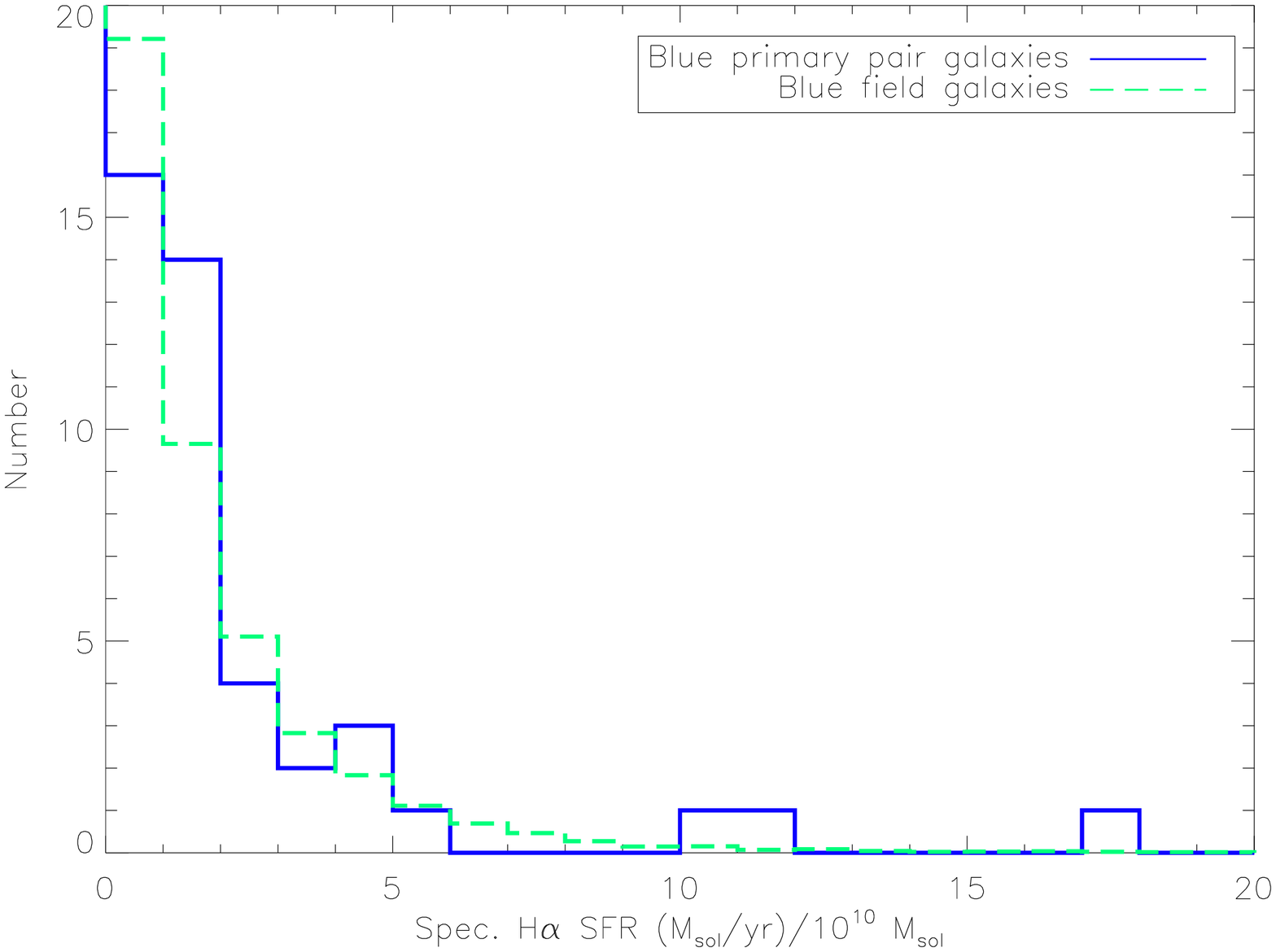}
  \includegraphics[width=2.3in,angle=90]{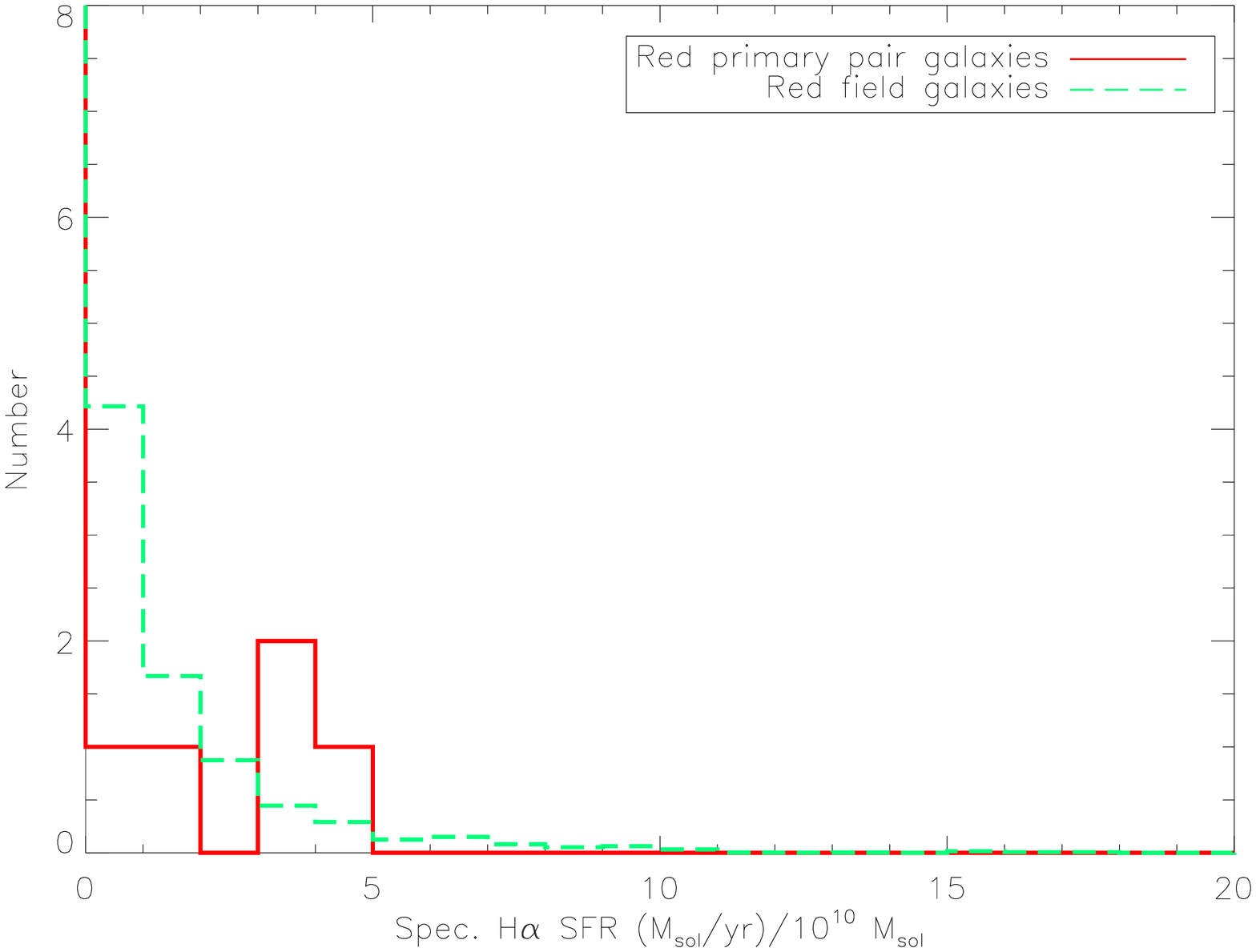}
\label{sb}
\caption{Distributions of SSFR for the blue primary galaxies (above left) 
	and the red primary galaxies (above right) in the
	 minor pairs sample, along with the average distribution of
	 SSFR for the matched set of blue field galaxies.}
\end{figure}

\begin{figure}[htb!]
\centering
  \includegraphics[width=2.3in,angle=90]{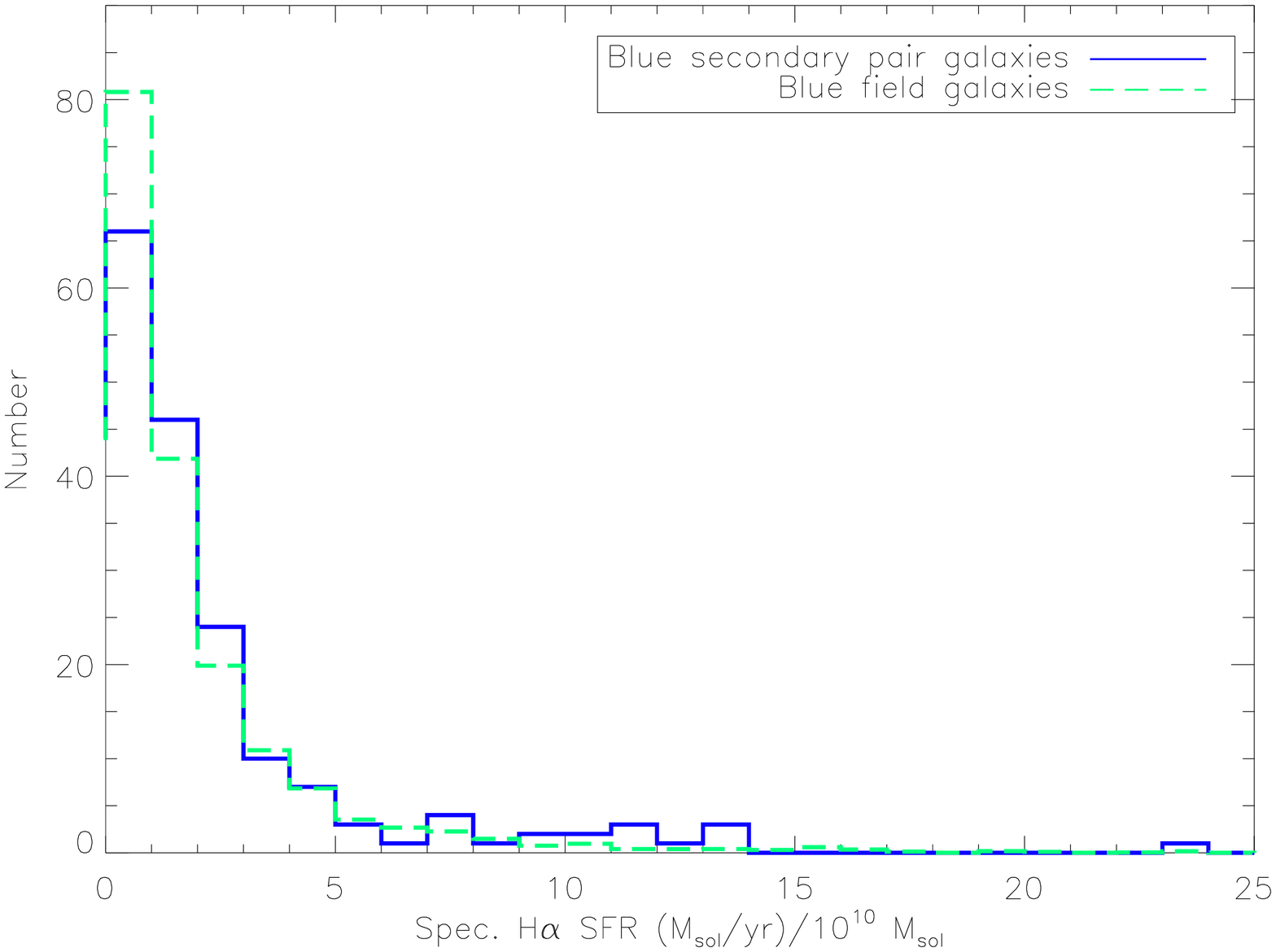}
  \includegraphics[width=2.3in,angle=90]{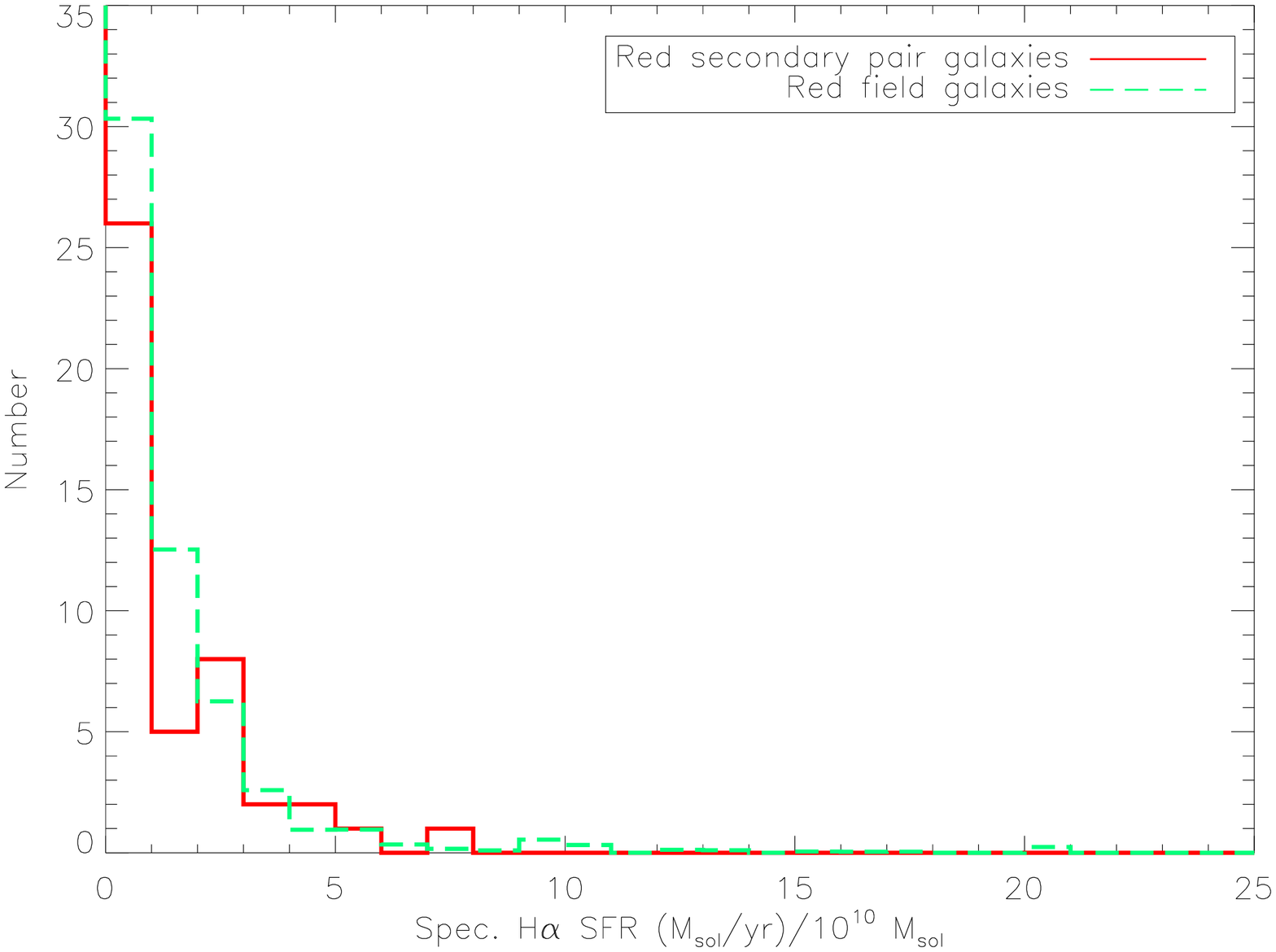}
\label{pb}
\caption{Distributions of SSFR for the blue secondary galaxies (above left) 
	and the red secondary galaxies (above right) in the
	 minor pairs sample, along with the average distribution of
	 SSFR for the matched set of blue field galaxies.}
\end{figure}

For the minor pairs sample, we find that the distributions of the SSFR 
of the primary or secondary blue starforming galaxies do not differ 
significantly from those of the matched subsets of field galaxies.
Figures~\ref{pb} and\ref{sb} show the SSFR distributions of the 
red and blue minor pair galaxies, and the average SSFR distributions of 
the random selection of field galaxies with matched absolute magnitude 
and color distributions.  A larger sample of pairs at small $\Delta D$ might 
yield a difference in the SSFR distributions because that is where the 
strongest signal is expected.  A problematic aspect of testing the minor 
pairs against a field sample is that minor companions should be very common.  
It is possible that many of our low luminosity field galaxies are also 
associated with minor interactions.  In contrast,
major interactions are less common, which may explain why we measure
differences in the SSFR distributions for the major pairs and matched
field galaxy samples.

Not all of the candidates for tidally triggered
star formation actually exhibit enhanced star formation. Some of
the apparent pairs may be superpositions and some of them may not yet
have reached perigalacticon.  Likewise, some of the galaxies with active 
star formation may not be reacting to the gravitational tidal 
interaction, but to supernova-enhanced star formation 
\citep{lada,elmegreen,boss} for the
less massive galaxies, or they may be interacting with another
companion hidden from our view on the far side of the galaxy.
The galaxy morphology strongly influences the strength or duration of 
the tidally triggered star formation \citet{mihos+hern96,tissera}.

\section{Specific Star Formation Rates and Pair Properties} \label{sfrprop}

Disparities in the SSFR distributions for major pair galaxies
and the matched field samples suggest systematic differences in 
star formation properties of galaxies in pairs and in the field, but
more conclusive evidence for tidally triggered star formation requires
a measured dependence on a parameter of the interaction.
We test the correlation between the specific star formation rate and the projected
spatial separation of the galaxy pairs.  

\citet{bgk} first showed that EW(H$\alpha$) and $\Delta D$ are correlated 
for major pair galaxies in a sample drawn from the CfA2 Redshift Survey.  
A number of other studies have also shown a link between star formation 
rates and pair
separation for major galaxy pairs (\citealp{lambas03,nikolic04,geller06}).
\citet{woods} examine the star formation properties of 57 galaxies in 
minor ($\left | \Delta m_r \right | > 2$) pairs, which do not show 
strong evidence for enhanced star formation rates.    Here we examine 
a much larger sample that enables study  of subsets of 
primary and secondary galaxies separately, and in subsets of color.

\begin{figure}[htb!]
\centering
 \includegraphics[width=2.3in,angle=90]{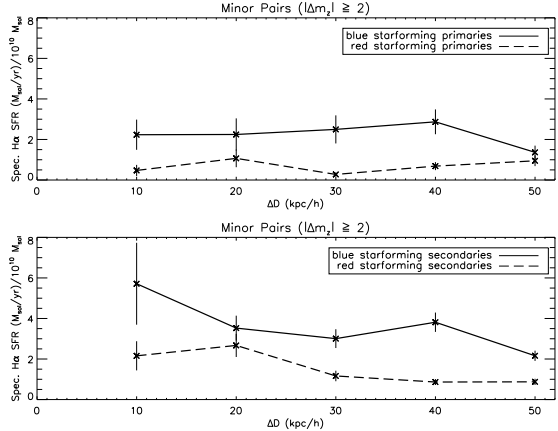}
\caption{The mean SSFR per $\Delta D$ bin for the red and blue starforming galaxies
	 in minor pairs.    The Spearman rank test detects a 
 	correlation between the individual SSFR and $\Delta D$ 
	 for the blue and for the red secondary starforming galaxies, but not
	for the blue or for the red primary starforming galaxies. }
\label{ssfr-dd-minor1}
\end{figure}

We find that the SSFR and $\Delta D$ are correlated in the secondary 
starforming galaxies in minor pairs, for both the blue and the red galaxy subsets. 
Table~\ref{correlations} gives a summary of the Spearman rank test 
correlation probabilities.  Figure~\ref{ssfr-dd-minor1} shows 
the mean SSFR vs. $\Delta D$ for the minor interactions.  
The correlations between SSFR and $\Delta D$ in the minor
secondary galaxies is detected even though
the sample is deficient in close pairs.  If our minor pairs sample 
had a flat distribution across $\Delta D$, as it is should in a 
complete sample, we would expect to measure stronger correlations.

 The primary starforming galaxies in minor pairs do not have a measurable 
correlation between SSFR and $\Delta D$ for either the blue or the red subsets.
We note that it may be particularly difficult to observe triggered
star formation in the minor primaries in our sample due to the deficiency 
in pairs at small $\Delta D$.  Our sample of minor primaries that
are classified as blue starforming galaxies comprises 68 galaxies,
and only 17 of those 68 galaxies have $\Delta D < 20$~kpc h$^{-1}$.
Even with a larger sample, we would expect the triggered star formation 
in the minor primaries to be weaker because galaxies with high luminosity 
tend to have lower gas contents.  Additionally, simulations of the Milky 
Way and LMC interaction demonstrate an asymmetry in the response to the 
gravitational interaction of the primary and secondary galaxies in  minor 
interactions, where the minor LMC is much more strongly affected than
the Milky Way \citep{mayer01,mastro},

\begin{table}[htb]
\caption{Spearman rank tests of SSFR versus $\Delta D$ correlation for
	      various samples.
\label{correlations} }
\begin{tabular}{llll}
\tableline \tableline
Sample & $P_{SR}$ & $C_{SR}$ & Sample Size \\
\tableline
Minor primary (blue SFG\tablenotemark{a}) & 0.46  & $-9.1\times10^{-2}$ & 68 \\
Minor primary (red SFG) & 0.68 & $ 6.6\times10^{-2}$ & 41 \\
Minor secondary (blue SFG) & $1.3\times10^{-2}$ & $-0.17$ & 220 \\
Minor secondary (red SFG) & $9.3\times10^{-3}$ & $-0.24$ & 118 \\
Minor primary \& secondary (blue \& red SFG) & $2.2\times10^{-3}$ & $-0.14$ & 447 \\
Major primary (blue SFG) & $2.1\times10^{-2}$ & $-0.16$ & 190 \\
Major primary (red SFG) & 0.53 & $-5.9\times10^{-2}$ & 118 \\
Major secondary (blue SFG) & $1.9\times10^{-3}$ & $-0.17$ & 335 \\
Major secondary (red SFG) & 0.20 & $-0.10$ & 167 \\
Major primary \& secondary (blue \& red SFG) & $4.1\times10^{-4}$ & $-0.12$ & 810\\
\tableline
\end{tabular}
\tablenotetext{a}{SFG: Star Forming Galaxy.}
\end{table}

\begin{figure}[htb!]
\centering
 \includegraphics[width=2.3in,angle=90]{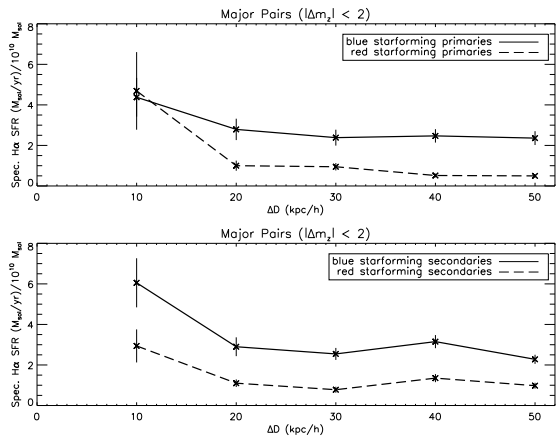}
\caption{The SSFR per $\Delta D$ bin for the starforming galaxies
	in the major pairs samples.  The Spearman rank test detects a 
 	correlation between the individual SSFR and $\Delta D$ 
	 for the blue primary starforming galaxies in the major pairs. }
\label{ssfr-dd-major1}
\end{figure}

The previous studies of major galaxy interactions that have demonstrated
a correlation between star formation indicators and projected 
separation  are not segregated by color.  Here, we
test the contribution to the SSFR-$\Delta D$ correlation by the
red and blue subsets of starforming galaxies.  We find that
the major galaxy pairs ($\left | \Delta m_z \right | < 2$) 
exhibit strong correlations between SSFR and $\Delta D$ for 
the blue galaxies (Table~\ref{correlations}).  This correlation 
holds for the blue primary and blue secondary galaxies in the 
major pairs when considered separately or together.  The subsets of 
red starforming galaxies  do not exhibit statistically significant 
correlations between SSFR and $\Delta D$, although the plot of 
mean SSFR versus $\Delta D$ appears to suggest a trend in high SSFR
at small $\Delta D$ (Figure~\ref{ssfr-dd-major1}).  However, when we 
measure the correlation between SSFR and $\Delta D$ for the entire 
set of starforming galaxies in the major
pairs sample, including both red and blue primaries and secondaries,
we do detect a very strong correlation between SSFR and $\Delta D$.
Our results suggest that the blue starforming galaxies dominate the
measurement of a correlation between formation indicators and projected 
separation in samples not separated by color.

The primary and secondary galaxies in major pairs exhibit symmetry 
in their response to the tidal interaction, in contrast to the minor 
pairs.  Our results agree with those of \citet{nikolic04}, who find that 
the relative $z$-band magnitude does not affect the star formation 
activity in their sample of SDSS pairs with 
$\left | \Delta m_z \right | < 2$.  The CfA2 Survey galaxy 
pairs with $\left | \Delta m_R \right | < 2$ in the study by
\citet{woods} also show similar distributions of EW($\alpha$) across
the range of small magnitude differences.  Both the brighter and
the fainter in the galaxy pairs with luminosity ratio $L_2/L_1 \leq 0.5$
($\left | \Delta m \right | < 0.75$) show similar increases in stellar 
birthrates compared to field samples in the 2dF field galaxy pairs of 
\citet{lambas03}; the galaxy pairs with $L_2/L_1 > 0.5$ show less star 
formation enhancement than pairs of similar luminosity.  Our results
are in general agreement with those of Lambas et al., except that we
find the secondary galaxies in $\left | \Delta m_z \right | \geq 2$
pairs have more significantly enhanced SSFRs than the primaries;
Lambas et al. find the opposite for in close ($\Delta D < 25$~kpc h$^{-1}$) 
pairs with $L_2/L_1 > 0.5$ ($\left | \Delta m \right | < 0.75$).  It
is unclear how the different distributions of relative magnitudes 
affect the comparison of our work with that of Lambas et al.

\subsection{Concentration Index and Pair Properties} \label{concentr}

As a complement to studying the star formation rates,
we also look for trends in the concentration index of the pair galaxies.
The concentration index, $C$, defined here as the ratio of the radii 
of apertures containing $90\%$ and $50\%$ of the $r$-band Petrosian 
magnitude, can be linked to galaxy type and color 
\citep[e.g. ][]{strateva01,kauffmann03}.  
Galaxies with concentration indexes in the range $2.0 < C < 2.6$ correspond to 
spiral and irregular types, while those with $C > 2.6$ usually 
correspond to early type galaxies \citep{strateva01}.  A high value of
$C$ indicates a relatively bright central region.  
\citet{nikolic04} discuss the use of the concentration index as a
morphological indicator for interacting galaxies.  They point out that 
late-type systems with strong central star formation would be wrongly
categorized.  

We avoid assigning a morphological type based on concentration index and
instead examine the correlation between the concentration index and $\Delta D$.  
The blue starforming secondary galaxies in the minor pairs exhibit a strong 
correlation between $C$ and $\Delta D$, the Spearman rank probability 
is $P_{SR} = 2.9 \times 10^{-5}$, where $C_{SR} = -0.28$.  Hence, galaxies 
at small separations tend to be have higher central concentrations of light.  
This result is consistent
with the SSFR-$\Delta D$ correlation, suggesting that a central burst
of star formation occurs in interacting galaxies, and is observed both in 
terms of a relatively bright central region and a high value of SSFR. 
 Similarly, we do not measure a correlation between $C$ and $\Delta D$ for
the  either the blue or red primary galaxies in the minor pairs.
The concentration index-$\Delta D$ correlation for pair galaxies
agrees well with the work of \citet{nikolic04}, who measure a
correlation between concentration index
and $\Delta D$ in their sample of SDSS close pairs.

\section{Specific Star Formation Rates and Galaxy Properties} \label{sfrmag}

In the previous section we compared the SSFRs for pair
samples with matched sets of field galaxies and found systematic
differences in their distributions in some cases.  
In this section we look more specifically at the SSFRs of pair galaxies 
as a function of $M_z$, compared to matched sets of field galaxies.  
Previous studies have yielded different conclusions on the role of
intrinsic luminosity in affecting star formation activity in pair 
galaxies.  \citet{lambas03} report no luminosity dependence for the 
mean star formation enhancement of their pair galaxies compared to 
representative field galaxies in their sample of close galaxy pairs 
($\Delta D < 25$~kpc h$^{-1}$ and $\Delta V < 100$~km s$^{-1}$) in the 2dF Survey.  
However, they find that a higher fraction of the more luminous
galaxies have a stellar birthrate parameter greater than the mean in 
their corresponding field sample.  

\begin{figure}[htb!]
\centering
   \includegraphics[width=2.3in,angle=90]{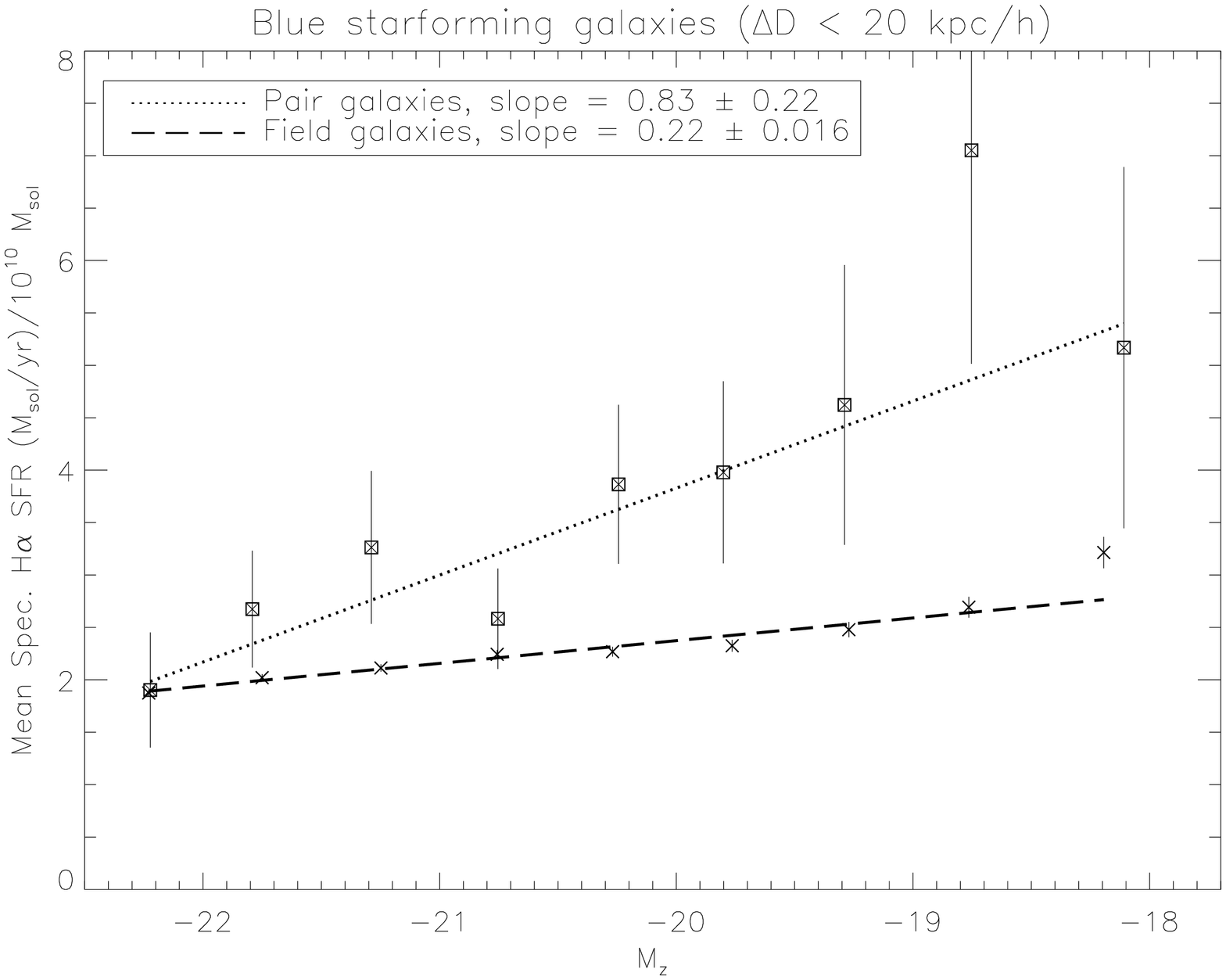}
   \includegraphics[width=2.3in,angle=90]{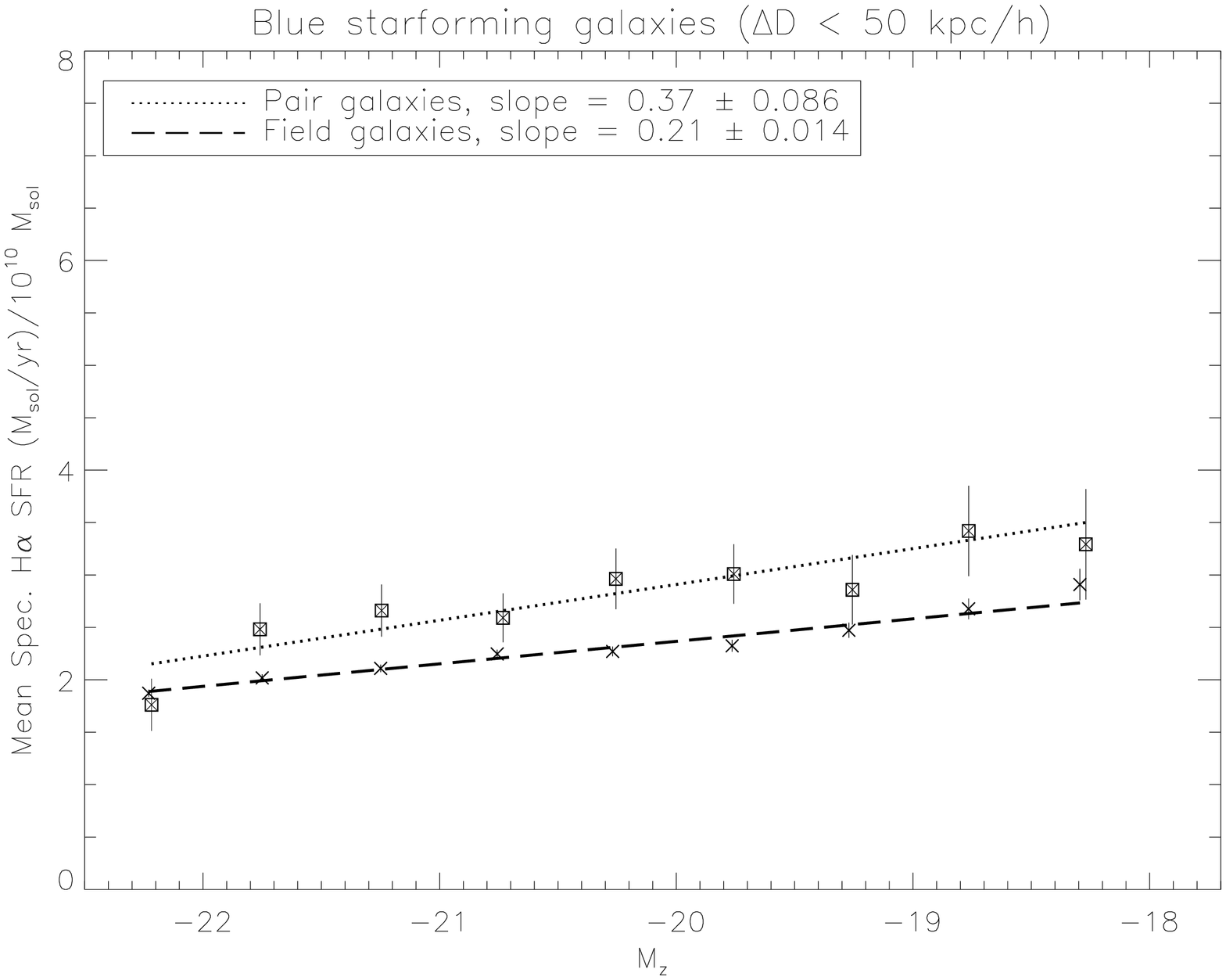}
\caption{The mean SSFR as a function of $M_z$ for the entire sample of blue 
	  starforming galaxies in major and minor pairs, and the mean SSFR 
	for the matched sets of field galaxies.  
	The plot on the left shows  the pairs with $\Delta D < 20$~kpc h$^{-1}$, 
	and the plot on the right shows the pairs with
       $\Delta D < 50$~kpc h$^{-1}$.  Although the galaxies in pairs have 
	greater mean SSFR at every $M_z$, the lower luminosity galaxies have 
	greater enhancement in their mean SSFRs compared to the field sample.}
\label{sfr-mag-dd20}
\end{figure}

Pair galaxies have a higher mean SSFR than field galaxies at every $M_z$
in our combined sample of blue starforming galaxies in major and minor
pairs (Figure~\ref{sfr-mag-dd20}).  This result holds for our entire
pairs sample with $\Delta D < 50$~kpc h$^{-1}$, and for the subset of pairs
with $\Delta D < 20$~kpc h$^{-1}$, where the tidally triggered star 
formation should be strongest.  We detect a very interesting
trend in the mean SSFR as a function of $M_z$: pair galaxies with lower 
intrinsic luminosity have a greater relative increase in mean SSFR than 
pair galaxies with higher intrinsic luminosity, compared to matched 
sets of field galaxies (Figure~\ref{sfr-mag-dd20}).  The relative
increase in mean SSFR for the  $\Delta D < 20$~kpc h$^{-1}$ pair galaxies 
is more pronounced than for the  $\Delta D < 50$~kpc h$^{-1}$ pair galaxies,
consistent with the idea that pairs at smaller $\Delta D$ being
more strongly affected by the tidally triggered star formation.
 For the  $\Delta D < 20$~kpc h$^{-1}$ pairs, the slopes  of the mean SSFR 
versus $M_z$ for the pairs and for the field galaxies differ by 
almost $4\sigma$.  This general trend is insensitive to the binning of 
the $M_z$ or the inclusion of the highest and lowest $M_z$ bins.

Our large dynamic range in $M_z$ ($-22.5 < M_z < -18$) contributes to our 
ability to detect the trend in increased SSFR as a function of $M_z$.
Our sample includes such a large dynamic range because the galaxies
in the minor pairs populate the extremes of the distribution
(see the distributions of $M_z$ in Figure~\ref{absz-m}).  Lambas et al. may 
not have detected the trend in enhanced star formation as a function of 
luminosity because of  their large uncertainty in the values of mean star 
formation excess compared to field galaxies (see Figure~8 in their paper).

The larger gas content that is expected in the lower luminosity pair galaxies 
compared to the more luminous pair galaxies probably contributes to their greater 
relative increase in SSFR over the matched sets of field galaxies.
Measuring the gas content of the galaxies would help to determine its role 
in the greater relative increase in SSFR for lower luminosity galaxies.
The relationship between relative increase in mean SSFR and $M_z$
may also be attributed to pair properties such as 
$\left | \Delta m_z \right |$ (a proxy for the mass ratio); the galaxies 
at the extremes of the
$M_z$ range are mostly minor pairs.  In \S\ref{sfrprop} we show that the
primary galaxies in minor pairs experience less triggered star formation
than the secondary galaxies in minor pairs.  A test of the
influence of the relative magnitude on the relationship between
increased SSFR and $M_z$ requires a pairs sample that is large enough
to measure mean SSFR for pairs compared to field galaxies for galaxy
pairs segregated by $\left | \Delta m_z \right |$ across a wide range 
in $M_z$.

\section{Conclusions} \label{conclusion}

We study a sample of 1204 galaxies in minor pairs
($\left | \Delta m_z \right | \geq 2$) and 2409 galaxies in
major pairs ($\left | \Delta m_z \right | < 2$) drawn from the
SDSS DR5.  We classify the galaxies in our pairs sample as
starforming galaxies, composites, or AGNs, and separate by color
to distinguish intrinsic galaxy properties from properties of the 
interaction. Our results suggest that gravitational tidal interactions trigger bursts
of star formation  in cases where the gravitational tidal force is 
relatively strong compared to the self-gravity of the galaxy: 
the secondary galaxy in minor pairs is more strongly affected by the 
interaction than the primary galaxy; both galaxies in major interactions
show enhanced star formation.  Blue galaxies appear to be more susceptible
to tidally triggered star formation than red galaxies, probably resulting
from their higher gas content \citep{barnes+hern96}.

In the sample of minor pairs, the secondary galaxies show 
a correlation between SSFR and $\Delta D$, for both subsets of blue and 
red starforming galaxies.  The primary galaxies in the minor pairs sample 
do not have measurable correlations between SSFR and $\Delta D$; however, 
this subset may be particularly disadvantaged by the lack of pairs at small 
separation.   Measurement of the correlation between the concentration index 
$C$ and $\Delta D$ for our minor pair galaxies supports our SSFR - $\Delta D$ 
results: the blue secondary galaxies show a correlation between high values 
of $C$ and small $\Delta D$; the red secondary galaxies do not.  Large 
$C$ corresponds to centrally bright galaxies, which may be experiencing a 
burst of central star formation.  

For the galaxies in major interactions, the subsets of blue
starforming primaries and blue starforming secondaries both show
a correlation between the SSFR and $\Delta D$, while the subsets of red
starforming galaxies do not.  A consistent explanation for these observations
is that the blue galaxies tend to be more gas-rich than the red 
galaxies.  When the galaxy experiences a gravitational tidal force
from its companion, the gas is driven to the center, where a burst
of star formation occurs.  The red galaxies, which tend to be
gas-poor, have little material available to form new stars, even if
they experience gravitational tidal forces \citep{barnes+hern96}.
Comparison of the distribution of SSFRs of the major pair galaxies with 
matched subsets of field galaxies reveal statistical differences for
the blue galaxies, again suggesting that the blue galaxies in major pairs
experience enhanced star formation rates.  

Galaxies in pairs have higher
mean SSFR at every absolute magnitude compared to matched sets of field
galaxies.  The relative increase in SSFR depends on the intrinsic luminosity 
of the galaxy; galaxies with lower intrinsic luminosities show a greater 
relative increase in mean SSFR compared to matched sets of field galaxies
than galaxies with high intrinsic luminosities do.  Our pairs sample is
the first to show this trend conclusively because of the 
large dynamic range in $M_z$ ($-22.5 < M_z < -18$).  Including both major 
and minor encounters allows us to obtain the large dynamic range because
the extremes of the range are populated mainly with minor pairs.

The greater relative increase in SSFR compared to the matched sets of field 
galaxies for the lower luminosity galaxies may be attributable to their larger
gas content compared to high luminosity galaxies, or it may be attributable
to pair properties such as $\left | \Delta m_z \right |$.  Measuring the
gas content of the galaxies is important for determining its role in the
greater relative increase in SSFR for lower luminosity galaxies.  It would
also be useful to obtain a pairs sample that is large enough to separate
into $\left | \Delta m_z \right |$ subsets across a wide range in $M_z$.

Our sample hints at triggered AGN activity in pair galaxies.
The fraction of AGN is greater in the pairs samples compared 
to matched sets of field galaxies with similar distributions 
of absolute magnitude and color.  We also find that the fraction of 
objects classified as AGN increases at small $\Delta D$ for the
primary galaxies in minor interactions.  The triggered AGN
observations are enticing but not conclusive: small aperture
spectroscopy for pairs at small apparent separations are needed.

A consistent picture on the effects of tidal triggering is emerging in
a variety of pairs samples.  Clarification of the outstanding issues
requires better sampling at small $\Delta D$ for a large dynamic
range of intrinsic luminosities, and measurements of the gas content
of the galaxies.  Comparisons with model predictions of observable
parameters would also contribute to a detailed understanding of
tidally triggered star formation and AGN activity.

\section{Acknowledgments}

The authors thank Elizabeth Barton, Lisa Kewley, and Daniel Fabricant 
for discussions that helped to improve this paper.  We thank Bill Wyatt 
for maintaining a local copy of the SDSS data products at the 
Harvard-Smithsonian Center for Astrophysics.  

Our research has made use of NASA's Astrophysics Data System 
Bibliographic Services.  This work is funded in part by the Smithsonian 
Institution.

    Funding for the SDSS and SDSS-II has been provided by the Alfred P. 
Sloan Foundation, the Participating Institutions, the National Science 
Foundation, the U.S. Department of Energy, the National Aeronautics and 
Space Administration, the Japanese Monbukagakusho, the Max Planck Society, 
and the Higher Education Funding Council for England. The SDSS Web Site 
is http://www.sdss.org/.

    The SDSS is managed by the Astrophysical Research Consortium for 
the Participating Institutions. The Participating Institutions are the 
American Museum of Natural History, Astrophysical Institute Potsdam, 
University of Basel, University of Cambridge, Case Western Reserve 
University, University of Chicago, Drexel University, Fermilab, the 
Institute for Advanced Study, the Japan Participation Group, Johns 
Hopkins University, the Joint Institute for Nuclear Astrophysics, 
the Kavli Institute for Particle Astrophysics and Cosmology, the 
Korean Scientist Group, the Chinese Academy of Sciences (LAMOST), 
Los Alamos National Laboratory, the Max-Planck-Institute for 
Astronomy (MPIA), the Max-Planck-Institute for Astrophysics (MPA), 
New Mexico State University, Ohio State University, University of 
Pittsburgh, University of Portsmouth, Princeton University, the 
United States Naval Observatory, and the University of Washington.

\end{document}